\documentclass[12pt]{article}
\usepackage{amssymb}
\usepackage{amsmath}
\usepackage{IEEEtrantools}
\usepackage{pdflscape}
\usepackage{afterpage}
\usepackage{capt-of}
\usepackage{graphicx}
\usepackage{enumitem}  

\numberwithin{equation}{section}

\newcommand{\be}{\begin{equation}}
\newcommand{\ee}{\end{equation}}
\newcommand{\non}{\nonumber}

\newcommand{\A}{\mathbb{A}}
\newcommand{\B}{\mathbb{B}}
\newcommand{\C}{\mathbb{C}}
\newcommand{\D}{\mathbb{D}}
\newcommand{\HH}{\mathbb{H}}
\newcommand{\K}{\mathbb{K}}
\newcommand{\M}{\mathbb{M}}

\newcommand{\R}{\mathbb{R}}
\newcommand{\T}{\mathbb{T}}
\newcommand{\U}{\mathbb{U}}
\newcommand{\tr}{\mathop{\rm tr}\nolimits}
\newcommand{\diag}{\mathop{\rm diag}\nolimits}

\usepackage[usenames,dvipsnames]{color}

\begin{document}

\begin{titlepage}
\strut\hfill UMTG--301
\vspace{.5in}
\begin{center}

\LARGE On Generalized $Q$-systems\\ 
\vspace{1in}
\large 
\large Zolt\'an Bajnok\footnote{
Wigner Research Centre for Physics, Konkoly-Thege Mikl\'os u. 29-33, 
1121 Budapest, Hungary,\\ bajnok.zoltan@wigner.mta.hu
}, Etienne Granet\footnote{Institut de Physique Th\'eorique, Paris 
Saclay, CEA, CNRS, 91191 Gif-sur-Yvette, France}${}^{,}$\footnote{Laboratoire 
de Physique de l’Ecole Normale Sup\'erieure, ENS, Universit\'e PSL, 
CNRS, Sorbonne Universit\'e, Universit\'e Paris-Diderot, Sorbonne 
Paris Cit\'e, Paris, France, etienne.granet@gmail.com }, Jesper Lykke Jacobsen${}^{2,3,}$\footnote{Sorbonne Universit\'e, \'Ecole Normale 
Sup\'erieure, CNRS, Laboratoire de Physique (LPENS), 75005 Paris, 
France, jesper.jacobsen@ens.fr}\\
and Rafael I. Nepomechie\footnote{
Physics Department,
P.O. Box 248046, University of Miami, Coral Gables, FL 33124 USA,
nepomechie@miami.edu}
\\[0.8in]
\end{center}

\vspace{.5in}

\begin{abstract}
We formulate $Q$-systems for the closed XXZ, open XXX and open
quantum-group-invariant XXZ quantum spin chains.  Polynomial solutions
of these $Q$-systems can be found efficiently, which in turn lead
directly to the admissible solutions of the corresponding Bethe ansatz
equations.
\end{abstract}

\end{titlepage}

\setcounter{footnote}{0}

\section{Introduction}\label{sec:intro}

An inconvenient truth about quantum integrable models --  well-known to 
experts but seldom acknowledged -- is that the corresponding Bethe ansatz (BA) 
equations (to which exact solutions of such models invariably reduce) are very 
difficult to solve. Various approaches to solving BA equations have 
been investigated, see e.g. \cite{Hagemans:2007, Hao:2013jqa} and 
references therein. Significant further progress on this problem was recently 
achieved in \cite{Marboe:2016yyn}, which formulated so-called 
$Q$-systems, whose polynomial solutions can be found efficiently;
the zeros of the fundamental $Q$-function are the sought-after Bethe 
roots. The $SU(2)$-invariant $Q$-system was an essential ingredient 
in the recent computation of torus partition functions 
\cite{Jiang:2017phk, Jacobsen:2018pjt}, which exploited also techniques 
from algebraic geometry.

The $Q$-systems in \cite{Marboe:2016yyn} were restricted to 
\emph{rational} BA equations for \emph{closed} spin chains with periodic 
boundary conditions. The purpose of this paper is to  
generalize the $SU(2)$-invariant $Q$-system  \cite{Marboe:2016yyn} 
in two different directions: from rational to trigonometric, and from 
closed to open. These new $Q$-systems will be used to compute 
partition functions for trigonometric vertex models and for vertex models with 
boundaries \cite{Bajnok:2019}.

The outline of this paper is as follows. In Sec. \ref{sec:closedXXX}, 
we review the $Q$-system from \cite{Marboe:2016yyn}
for the closed XXX spin chain with periodic 
boundary conditions. However, we provide an 
alternative derivation based on \cite{Granet:2019}, which is convenient for deriving generalizations.
In Sec. \ref{sec:closedXXZ}, we formulate a $Q$-system for the 
closed XXZ spin chain with periodic boundary conditions. We then turn 
to open spin chains. In Sec. \ref{sec:openXXX}, we formulate a $Q$-system
for the $SU(2)$-invariant open XXX spin chain. A $Q$-system
for the quantum-group-invariant open XXZ spin chain 
\cite{Pasquier:1989kd} is formulated in Sec. \ref{sec:openXXZ}.
We conclude in Sec. \ref{sec:conclusion} with a brief summary and a 
list of some interesting open problems.

\section{Closed XXX $Q$-system}\label{sec:closedXXX}

In this section we review the $Q$-system \cite{Marboe:2016yyn} whose polynomial solutions
provide the full spectrum of the closed XXX spin chain of length $N$ with periodic 
boundary conditions, whose Hamiltonian is given by
\be
\HH = \sum_{k=1}^{N} \vec \sigma_{k} \cdot \vec \sigma_{k+1} \,, 
\qquad \vec\sigma_{N+1} \equiv \vec\sigma_{1}  \,.
\label{eq:HclosedXXX}
\ee
First the model is introduced and its solution by the algebraic BA method is
recalled, together with the physicality conditions for the Bethe roots.
We then describe the relevant $Q$-system and construct its solution
explicitly. We show that polynomial solutions are in one-to-one correspondence
with the physical solutions. 

\subsection{Review of the algebraic BA solution}

The closed spin-1/2 XXX spin chain and its solution can be succinctly 
formulated with the help of an $SU(2)$-invariant 
solution of the Yang-Baxter equation given by the $4 \times 4$ $R$-matrix (see 
e.g. \cite{Faddeev:1996iy})
\begin{equation}
\mathbb{R}(u)=(u-\frac{i}{2})\mathbb{I}+i\mathbb{P}\,,
\label{eq:Rmatrix}
\end{equation}
where $\mathbb{P}$ is the permutation matrix, $\mathbb{I}$ is the 
identity matrix, and $u$ is the spectral
parameter. For $N$ sites with periodic boundary conditions, one can
introduce the monodromy matrix $\mathbb{M}$ and the transfer matrix
$\mathbb{T}$ as 
\begin{equation}
\mathbb{T}(u)=\tr_{0}(\mathbb{M}_{0}(u))\,, \qquad
\mathbb{M}_{0}(u)=\mathbb{R}_{01}(u)\, \mathbb{R}_{02}(u)\dots\mathbb{R}_{0N}(u) \,.
\label{eq:MandT}
\end{equation}
An auxiliary space denoted by index $0$ has been introduced, and lower
indices help indicate the spaces in which operators act. The transfer 
matrix is obtained by tracing over the auxiliary space, and thus acts on the
quantum space, which is the $N$-fold tensor product of $\mathbb{C}^{2}$
accommodating all possible states of spin-up and spin-down. As a consequence of 
the Yang-Baxter equation, the transfer matrix
forms a one-parameter family of commuting operators
\begin{equation}
\left[\mathbb{T}(u),\mathbb{T}(v)\right]=0\,,
\label{eq:commuting}
\end{equation}
and generates conserved charges in involution, including the Hamiltonian
of the system (\ref{eq:HclosedXXX}). 

We are interested in the eigenvectors and eigenvalues of the transfer
matrix. The former can be generated from the all spin-up reference state
\be
|0\rangle = {1\choose 0}^{\otimes N} 
\label{reference}
\ee
by acting with a matrix element of the monodromy
matrix as
\begin{equation}
\mathbb{B}(u_{1})\dots\mathbb{B}(u_{M})\vert0\rangle\equiv\vert 
u_{1},\dots,u_{M}\rangle\,,
\qquad\mathbb{M}_{0}(u) =\text{\ensuremath{\left(\begin{array}{cc}
 \mathbb{A}(u)  &  \mathbb{B}(u)\\
 \mathbb{C}(u)  &  \mathbb{D}(u) 
\end{array}\right)}}\,.
\label{eq:Bethevector}
\end{equation}
The eigenvalues $T(u)$ of the transfer matrix 
\begin{equation}
\mathbb{T}(u)\vert u_{1},\dots,u_{M}\rangle=T(u)\vert u_{1},\dots,u_{M}\rangle
\end{equation}
satisfy the $TQ$-relation
\begin{equation}
T(u)\, Q(u)=(u+\frac{i}{2})^{N}Q(u-i)+(u-\frac{i}{2})^{N}Q(u+i)\,,
\label{eq:TQ}
\end{equation}
where $Q$ encodes the Bethe roots $\{u_{i}\}$: 
\begin{equation}
Q(u)=\prod_{j=1}^{M}(u-u_{j}) \,.
\end{equation}
As follows from the definition of the transfer matrix and
its commutativity property (\ref{eq:commuting}), $T(u)$ is a polynomial in $u$, and is thus
regular at $u_{j}$.
The $TQ$-relation (\ref{eq:TQ}) then
leads to the BA equations for the roots:
\begin{equation}
\left(\frac{u_{j}+\frac{i}{2}}{u_{j}-\frac{i}{2}}\right)^{N}=
-\prod_{k=1}^{M}\frac{u_{j}-u_{k}+i}{u_{j}-u_{k}-i} \,, \qquad j = 1, 
\ldots, M \,.
\label{eq:BA}
\end{equation}
For roots with multiplicities, we have further equations 
\cite{Izergin:1982hy, Avdeev:1985cx}. Since repeated
roots do not seem to appear in this model (see e.g. \cite{Hao:2013jqa}), 
we assume that roots never coincide. 

We call a solution of the BA equations \emph{physical} if the corresponding 
Bethe vector (\ref{eq:Bethevector}) is an eigenvector of the transfer 
matrix.
Unfortunately, not all solutions of the BA equations are physical.
Solutions that contain the roots $\pm\frac{i}{2}$ can be unphysical, i.e. they might
solve the BA equations, but there is no related eigenvector of the
transfer matrix. We define a solution $\{u_{1},\dots,u_{M}\}$ of
the BA equations to be \emph{admissible}, if all roots are finite and
pairwise distinct; and, if they are of the form 
$\{\frac{i}{2},-\frac{i}{2},u_{1},\dots,u_{M-2}\}$ (which we call a 
\emph{singular} solution), then the further constraint
\begin{equation}
\prod_{j=1}^{M-2}\frac{(u_{j}+\frac{i}{2})}{(u_{j}-\frac{i}{2})}
\frac{(u_{j}+\frac{3i}{2})}{(u_{j}-\frac{3i}{2})}=(-1)^{N} 
\label{eq:physicality}
\end{equation}
is satisfied. It was shown in \cite{Nepomechie:2013mua} that 
admissibility implies physicality, and the converse follows from 
Lemmata 2 and 4 of \cite{Granet:2019}. Hence, admissibility and 
physicality are equivalent. The number ${\cal 
N}(N,M)$ of admissible solutions of the BA equations with $M\le N/2$ 
has been conjectured to be given by (see e.g. \cite{Hao:2013jqa})
\begin{equation}
	{\cal N}(N,M) = {N \choose M} - {N \choose M-1} \,.
	\label{expectedXXX}
\end{equation}	

Alternatively, it was observed in \cite{Marboe:2016yyn} that the 
polynomial solutions of a $Q$-system on an appropriately chosen diagram can be computed efficiently,
and correctly account for the physical solutions. In the remainder of 
this section, we provide an alternative derivation of these results based on \cite{Granet:2019}, which we will 
subsequently use to generalize this $Q$-system.

\subsection{$Q$-system}

\begin{figure}[htb]
\centering
	\includegraphics[width=12cm]{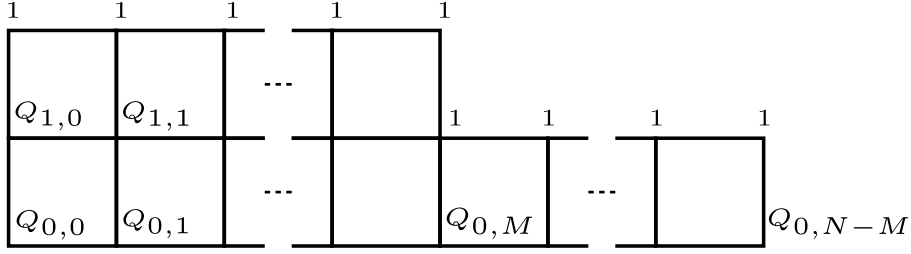}
\caption{Non-trivial $Q$-functions for the closed XXX spin chain.}
\label{QQdiagramXXX}
\end{figure}

For given values of $N$ and $M$, the $Q$-functions $Q_{a,s}$ are 
defined on a Young
tableau with the indices referring to the vertex $(a,s)$, 
where the $a$-axis is vertical and the $s$-axis is horizontal,
see Fig. \ref{QQdiagramXXX}. These $Q$-functions satisfy the $QQ$-equations,
which are formulated around a face as \footnote{A 
$Q$-function is defined up to a multiplicative constant. For 
definiteness, we generally treat $\propto$ as equality, or (as in 
(\ref{eq:Qprime})) with an extra minus sign.}
\begin{equation}
Q_{a+1,s}(u)\, Q_{a,s+1}(u)\propto Q_{a+1,s+1}^{+}(u)\, Q_{a,s}^{-}(u)
-Q_{a+1,s+1}^{-}(u)\, Q_{a,s}^{+}(u)\,,
\label{eq:QQXXX}
\end{equation}
where $f^{\pm}(u)=f(u\pm\frac{i}{2})$. The relevant diagram for the
closed XXX spin chain with the boundary conditions, $Q_{2,s}=1$,
$Q_{1,s\ge M}=1$, is displayed in Fig. \ref{QQdiagramXXX}. The initial
condition 
\begin{equation}
Q_{0,0}(u)=u^{N}\,, \qquad Q_{1,0}(u)=Q(u)=\prod_{j=1}^{M}(u-u_{j}) 
\,, \label{eq:QQXXXinit}
\end{equation}
leads to a unique solution of the $Q$-system. The degree of the 
polynomial $Q_{a,s}(u)$ is given by the number of boxes in the Young tableau
to the right and top of the vertex $(a,s)$.
Let us see how we can proceed column-by-column and express all $Q$-functions in terms
of $Q_{0,0}$ and $Q_{1,0}$. 

The $QQ$-equation for $(a,s)=(1,0)$ can be solved easily
\begin{equation}
Q_{1,1}(u)=Q_{1,0}^{+}(u)-Q_{1,0}^{-}(u)\equiv Q'_{1,0}(u)=Q'(u)\,,
\label{eq:Qprime}
\end{equation}
where we have introduced the discrete derivative defined by 
\begin{equation}
	f'(u)=f^{+}(u)-f^{-}(u) \,.
\label{eq:discretederiv}
\end{equation}	
The function $Q_{1,1}$ is automatically a polynomial of degree $M-1$.
The equation for $(a,s)=(0,0)$ gives 
\begin{align}
Q_{0,1}Q_{1,0} & =Q_{0,0}^{-}Q_{1,1}^{+}-Q_{0,0}^{+}Q_{1,1}^{-} \,.
\label{eq:as00}
\end{align}
Making use of (\ref{eq:Qprime}) and (\ref{eq:as00}), it follows that 
\begin{equation}
Q_{0,1}Q=Q_{0,0}^{-}Q^{++}+Q_{0,0}^{+}Q^{--}-Q(Q_{0,0}^{-}+Q_{0,0}^{+})\,,
\label{eq:Q01Q}
\end{equation}
or 
\begin{equation}
(Q_{0,1}+Q_{0,0}^{-}+Q_{0,0}^{+})Q = Q_{0,0}^{-}Q^{++}+Q_{0,0}^{+}Q^{--}\,.
\end{equation}
Recognizing the RHS of the above equation 
as the RHS of the $TQ$-relation (\ref{eq:TQ}), one 
obtains
\begin{equation}
T=Q_{0,1}+Q_{0,0}^{-}+Q_{0,0}^{+} \,.
\label{eq:TfromQ01}
\end{equation}
Polynomiality of $Q_{0,1}$ is equivalent to the polynomiality of
$T$, which leads to the BA equations (\ref{eq:BA}). 

\subsubsection{$Q_{a,s}$ in terms of $Q$ and $P$}

We now show that the polynomiality of the remaining $Q$-functions
is equivalent to the admissibility of $\{u_{1},\dots,u_{M}\}$. To 
this end, we define a function $P(u)$, such that 
\begin{equation}
Q_{0,0}=P^{+}Q^{-}-P^{-}Q^{+} \,.
\label{eq:PQ}
\end{equation}
Using this parametrization for $Q_{0,0}$, one can easily show that
\begin{equation}
Q_{0,1}\propto P'^{+}Q'^{-}-P'^{-}Q'^{+}\,,
\label{eq:Q01}
\end{equation}
where prime denotes discrete derivative (\ref{eq:discretederiv}).
Repeating the calculations starting from $(a,s)=(1,1)$ and 
$(a,s)=(0,1)$,
we arrive at 
\begin{equation}
Q_{1,2}=Q''\,, \qquad Q_{0,2} \propto P''^{+}Q''^{-}-P''^{-}Q''^{+} \,.
\end{equation}
This can be iterated further
\begin{equation}
Q_{1,n}=Q^{(n)}\,, \qquad Q_{0,n} \propto P^{(n)+}Q^{(n)-}-P^{(n)-}Q^{(n)+} \,,
\label{eq:QQsltnXXX}
\end{equation}
where the superscript $(n)$ denotes the $n^{th}$ discrete derivative. In short, all
$Q$-functions can be expressed in terms of $P$ and $Q$. Clearly,
if $P$ is a polynomial, then all $Q$-functions are polynomial. In
the following we show that the polynomiality of $P$ is in fact equivalent
to the polynomiality of $Q_{0,2}$. We also derive that polynomiality
of $P$ is equivalent to the admissibility of the roots $\{u_{1},\dots,u_{M}\}$. 

\subsubsection{Construction of $P$}

We construct $P$ as in \cite{Granet:2019} by generalizing the approach in \cite{Pronko:1998xa} 
(which implicitly assumes that all Bethe roots are regular)
to the case of a singular solution. String configurations have roots that differ by $i$: 
$u_{i_{1}}-u_{i_{2}}=i$,
and it is well known (see e.g. \cite{Granet:2019}) that the only exact 
string solution consists of one pair of singular roots
$u_{1}=\frac{i}{2}$ and $u_{2}=-\frac{i}{2}$. In the presence of such
singular roots, the $Q$-function takes the following form
\begin{equation}
Q(u)=u^{+}u^{-}\bar{Q}(u)\,, \qquad\bar{Q}(u)=\prod_{u_{j}\neq\pm\frac{i}{2}}(u-u_{j}) \,.
\label{eq:barQ}
\end{equation}
We start by dividing (\ref{eq:PQ}) by $Q^{+}Q^{-}$. We need
to write 
\begin{equation}
R(u)=\frac{u^{N}}{Q^{+}Q^{-}}=\frac{u^{N-2}}{u^{++}u^{--}\bar{Q}^{+}\bar{Q}^{-}}
\label{eq:R1}
\end{equation}
in the form 
\begin{equation}
R=\frac{P^{+}}{Q^{+}}-\frac{P^{-}}{Q^{-}}=\left(\frac{P}{Q}\right)^{'} \,,
\label{eq:Rdiff}
\end{equation}
i.e. we need to ``integrate'' $R$ in the discrete sense. To this end, we
perform a partial fraction decomposition of (\ref{eq:R1})
\begin{equation}
R=\pi+\frac{q_{+}}{\bar{Q}^{+}}+\frac{q_{-}}{\bar{Q}^{-}}+\frac{a_{+}}{u^{++}}+\frac{a_{-}}{u^{--}}\,,
\label{eq:Rdecomp}
\end{equation}
where $\pi$ is a polynomial of order $N-2M$, the polynomials $q_{\pm}$
have degree less than $\bar{Q}$, while $a_{\pm}$ are constants.
Using the relation
\begin{equation}
R^{+}+R^{-}=\frac{T}{Q^{++}Q^{--}}=\frac{T}{u^{+}u^{-}u^{+++}u^{---}\bar{Q}^{++}\bar{Q}^{--}}\label{eq:RpR}
\end{equation}
which follows from the $TQ$-relation (\ref{eq:TQ}), one can investigate all the
singularities explicitly. In particular, the RHS of (\ref{eq:RpR})
has no singularities at the zeros of $\bar{Q}$, implying
\begin{equation}
q_{+}=q^{+}\,, \qquad q_{-}=-q^{-}\,,
\end{equation}
for some polynomial $q(u)$. The coefficients $a_{\pm}$ can be determined
from the residues of (\ref{eq:RpR}) at $u=\mp\frac{i}{2}$: 
\begin{equation}
a_{\pm}=\mp\frac{T(\mp\frac{i}{2})}{2i \bar{Q}(\pm\frac{i}{2})\, \bar{Q}(\mp\frac{3i}{2})} \,.
\label{eq:apm}
\end{equation}
The polynomial $\pi$ can always be written as
\begin{equation}
\pi=\rho'=\rho^{+}-\rho^{-} \,,
\end{equation}
where $\rho$ is a polynomial.
Clearly, $\rho$ is defined up to a constant. Changing this constant
by $a$ modifies $P$ as $P+aQ$. This additional term, however,
disappears from $R$ and $Q_{0,0}$, thus is irrelevant for us. 

In the absence of singular roots, we have $\bar{Q}=Q$ and 
$a_{\pm}=0$; hence, the polynomial
$P=\rho Q+q$ satisfies (\ref{eq:Rdiff}), which implies the required
Eq. (\ref{eq:PQ}), see also \cite{Pronko:1998xa}. In the presence of 
singular roots,
the ``integration'' of $R$ in (\ref{eq:Rdiff}) requires to ``integrate''
$u^{-1}$, appearing in (\ref{eq:Rdecomp}). To this end, we define the 
function $p(u)$ by
\begin{equation}
p'(u)=\frac{1}{u}\,, \qquad 
p(u)=-i\psi(-iu+\frac{1}{2})\,, 
\label{pfuncdef}
\end{equation}
where $\psi(u)$ is the digamma function
\begin{equation}
\psi(u)=-\gamma+\sum_{n=0}^{\infty}\left(\frac{1}{n+1}-\frac{1}{n+u}\right) \,.	
\end{equation}
In view of the fact
\begin{align}
\frac{a_{+}}{u^{++}}+\frac{a_{-}}{u^{--}} & = \left( a_{+} p^{++} + 
a_{-} p^{--} \right)' 
= \left[\frac{1}{2}(a_{+} - a_{-})(p^{++} - p^{--}) + 
\frac{1}{2}(a_{+} + a_{-})(p^{++} + p^{--}) \right]' \non \\
& = \left[\frac{1}{2}(a_{+} - a_{-})(\frac{1}{u^{+}} + \frac{1}{u^{-}}) + 
\frac{1}{2}(a_{+} + a_{-})(p^{++} + p^{--}) \right]' \,,
\end{align}
we see that the function $P$ satisfying (\ref{eq:Rdiff}) takes the
form
\begin{equation}
P=\rho 
Q+u^{+}u^{-}q+(a_{+}-a_{-})u\bar{Q}+\frac{1}{2}(a_{+}+a_{-})(p^{++}+p^{--})Q \,.
\label{eq:Presult}
\end{equation}
 It is a polynomial if and only if $a_{+}=-a_{-}$, i.e. when
\begin{equation}
(-1)^{M}\frac{\bar{Q}(+\frac{i}{2})\bar{Q}(+\frac{3i}{2})}{\bar{Q}(-\frac{i}{2})\bar{Q}(-\frac{3i}{2})}=1
\end{equation}
is satisfied. Here we used the $TQ$-relation (\ref{eq:TQ}) to eliminate $T(\pm\frac{i}{2})$
in (\ref{eq:apm}).
Clearly this is the admissibility condition for singular solutions
(\ref{eq:physicality}). Thus, we have just proven that polynomiality
of $P$ is equivalent to the admissibility of the roots. 

Even if $P$ is not a polynomial, the relation (\ref{eq:Rdiff}) implies
(\ref{eq:PQ}), which leads to 
\begin{equation}
T=P^{++}Q^{--}-P^{--}Q^{++}\,.
\end{equation}
This implies that $P$ also satisfies the $TQ$-relation
\begin{equation}
TP=(u^{+})^{N}P^{--}+(u^{-})^{N}P^{++}\,.
\end{equation}
Thus $P$ and $Q$ are the two independent solutions of this second
order difference equation, and (\ref{eq:PQ}) is the corresponding 
Wronskian relation. 
It has been known (see e.g. \cite{Mukhin:2009, Tarasov:2018}) that
the two independent solutions of the $TQ$-relation
are both polynomial iff the Bethe state (\ref{eq:Bethevector}) 
is an eigenstate of the transfer matrix. 

Finally, let us investigate the polynomiality of
\begin{equation}
Q_{0,2}=Q_{0,1}^{+}+Q_{0,1}^{-}+P'^{++}Q'^{--}-P'^{--}Q'^{++}\,.
\end{equation}
Since $Q_{0,1}$ is a polynomial\footnote{Recall from (\ref{eq:TfromQ01})
that polynomiality of 
$Q_{0,1}$ is equivalent to polynomiality of $T$; and the latter is 
evident from
\begin{equation}
	T = (u^{+})^{N-1} u^{---} \frac{\bar{Q}^{--}}{\bar{Q}} + 
	(u^{-})^{N-1} u^{+++} \frac{\bar{Q}^{++}}{\bar{Q}} \,, \non
\end{equation}
which follows from the $TQ$-relation (\ref{eq:TQ}) and 
(\ref{eq:barQ}), and which has vanishing residues at the zeros of 
$\bar{Q}$ by virtue of the BA equations.}, 
we investigate the regularity of
the remaining part at $u=0$. Since $Q(\pm\frac{i}{2})=0$ we can
see that
\begin{equation}
Q'^{--}(0)=-Q(-\frac{3i}{2})\,, \qquad Q'^{++}(0)=Q(\frac{3i}{2}) \,,
\end{equation}
which are not zero. We now focus on the pole contributions at $u=0$.
They can only come from the terms proportional to $p(u)$, which have
poles at $u=-i(n+\frac{1}{2})$ for any integer $n\geq0$, with 
residues $-1$. Thus the singular
parts can arise as
\begin{align}
P'^{++}(\epsilon) & =P(\frac{3i}{2}+\epsilon)+\dots
=\frac{1}{2}(a_{+}+a_{-}) 
Q(\frac{3i}{2})(p^{+++++}(\epsilon)+p^{+}(\epsilon))=0+\dots \non \\
P'^{--}(\epsilon) & =-P(-\frac{3i}{2}+\epsilon)+\dots
=-\frac{1}{2}(a_{+}+a_{-})Q(-\frac{3i}{2})(p^{-----}(\epsilon)+p^{-}(\epsilon)) \non\\
& = \frac{1}{\epsilon}(a_{+}+a_{-}) Q(-\frac{3i}{2})+\dots \,,
\end{align}
where we have omitted regular terms in $\epsilon.$ The singular part
of $Q_{0,2}(\epsilon)$ is then $-\frac{1}{\epsilon}(a_{+}+a_{-})Q(\frac{3i}{2})Q(-\frac{3i}{2})$
whose vanishing implies $a_{+}=-a_{-}$, i.e. the polynomiality of
$P$.

We can thus conclude that the following four properties are equivalent: 
\begin{enumerate}[label=(\roman*)]
\item $P$ is a polynomial
\item all $Q_{a,s}$-functions are polynomial
\item the roots $\{u_{1},\dots,u_{M}\}$ are admissible solutions of the
BA equations
\item the Bethe vector is an eigenvector of the transfer matrix 
\end{enumerate}

\subsubsection{An example}\label{sec:exampleXXXclosed}

We conclude this subsection with an elementary explicit example of how to
use the Q-system (\ref{eq:QQXXX})-(\ref{eq:QQXXXinit}) to obtain all
the Bethe roots for modest values of $N\equiv L$ and $M$ using
a simple-minded implementation in {\tt Mathematica}. 
(A more general and sophisticated code is provided in \cite{Marboe:2016yyn}.)
We begin by defining the functions $Q_{0,0}(u)$ and $Q_{1,0}(u)$ 
\begin{verbatim}
Q[0, 0, u_] := u^L;  
Q[1, 0, u_] := Sum[c[k] u^k, {k, 0, M-1}] + u^M;
\end{verbatim}
where the coefficients $c_{k}$ are to be determined.
We also define the functions $Q_{1,n}(u)$ and $Q_{0,n}(u)$ 
by\footnote{The equation for $Q_{1,n}$ is valid only for 
$1\le n \le M$. For $n=M$, we find $Q_{1,M} \propto 1$.}
\begin{verbatim}
Q[1, n_, u_] := Q[1, n-1, u + I/2] - Q[1, n-1, u - I/2];
Q[0, n_, u_] := (Q[1, n, u + I/2] Q[0, n-1, u - I/2] - 
   Q[1, n, u - I/2] Q[0, n-1, u + I/2])/ Q[1, n-1, u];
\end{verbatim}
which follow from (\ref{eq:QQXXX}) with $a=1$ (setting $Q_{2,s}=1$) 
and $a=0$, respectively.
The functions $Q_{1,n}(u)$ are evidently polynomials in $u$; the key point 
is that the functions $Q_{0,n}(u)$ must also be polynomials in $u$. In order to 
ensure the latter requirement (the so-called zero-remainder conditions), 
we first use the built-in symbol {\tt PolynomialRemainder} to define the polynomials $y_{n}(u)$
\begin{verbatim}
y[n_, u_] := PolynomialRemainder[Numerator[Together[Q[0, n, u]]], 
  Denominator[Together[Q[0, n, u]]], u];
\end{verbatim}
We then solve for the coefficients $c_{k}$ that make $\{ y_{1}(u), 
\ldots, y_{M}(u) \}$ vanish for all values of $u$\footnote{For 
some examples, the coefficients $c_{k}$ can be determined using fewer
than $M$ equations.}
\begin{verbatim}
sol = Solve[Table[CoefficientList[y[n, u], u] == 0, {n, 1, M}], 
	Table[c[k], {k, 0, M-1}]]
\end{verbatim}

Let us consider as an example the case $L=6, M=2$. The above code 
generates 9 solutions, in agreement with (\ref{expectedXXX}).
For each of these solutions, one can obtain the corresponding Bethe roots
by solving for the zeros of $Q_{1,0}(u)$. For example, for the first solution
\begin{verbatim}
	Solve[(Q[1, 0, u] /. sol[[1]]) == 0, u] // Flatten
\end{verbatim}
we obtain the Bethe roots $\pm i/2$. 

\section{Closed XXZ $Q$-system}\label{sec:closedXXZ}

In this section we present a generalization of the $Q$-system
for the closed XXZ spin chain of length $N$ with periodic boundary conditions, 
whose Hamiltonian is given by
\be
\HH = \sum_{k=1}^{N} \left[ \sigma^x_k \sigma^x_{k+1} +
\sigma^y_k \sigma^y_{k+1} + {1\over 2}( q + q^{-1}) \sigma^z_k
\sigma^z_{k+1}
\right]\,, \qquad \vec\sigma_{N+1} \equiv \vec\sigma_{1} \,.
\label{eq:HclosedXXZ}
\ee
We show that there is a notion of polynomial
solutions of the $QQ$-equations, which determine the spectrum of the
closed XXZ model. 

\subsection{Review of the algebraic BA solution}

The XXZ spin chain is related to the trigonometric generalization
of the rational $R$-matrix (\ref{eq:Rmatrix}): \footnote{The XXX limit 
can be recovered by the rescalings $u\to\epsilon u$,
$\eta\to i\epsilon$ and by then taking the $\epsilon\to0$ limit.}
\begin{equation}
\mathbb{R}(u)=\left(\begin{array}{cccc}
\sinh(u+\frac{\eta}{2}) & 0 & 0 & 0\\
0 & \sinh(u-\frac{\eta}{2}) & \sinh(\eta) & 0\\
0 & \sinh(\eta) & \sinh(u-\frac{\eta}{2}) & 0\\
0 & 0 & 0 & \sinh(u+\frac{\eta}{2}) 
\end{array}\right) \,.
\label{eq:RXXZ}
\end{equation}
The monodromy and transfer matrices can be introduced
by the analogous formulae to the XXX case (\ref{eq:MandT}). The off-diagonal
elements of the monodromy matrix (\ref{eq:Bethevector}) can be used
as creation and annihilation operators. The $\mathbb{B}$ operators,
by acting on the all spin-up reference state (\ref{reference}), create eigenstates of
the transfer matrix. The eigenvalue $T(u)$ of the transfer matrix 
$\mathbb{T}(u)$ satisfies the $TQ$-relation
\begin{equation}
T(u)\, Q(u)=\sinh^{N}(u+\tfrac{\eta}{2})\, 
Q(u-\eta)+\sinh^{N}(u-\tfrac{\eta}{2})\, Q(u+\eta) \,,
\end{equation}
where now 
\begin{equation}
Q(u)=\prod_{j=1}^{M}\sinh(u-u_{j}) \,.
\end{equation}
Alternatively, we can switch to the variable $t=e^{u}$. By construction,
$T(u)$ is a polynomial of $t$ and $t^{-1}$, regular at $t_{j}=e^{u_{j}}$,
which implies the BA equations
\begin{equation}
\left(\frac{\sinh(u_{k}+\frac{\eta}{2})}{\sinh(u_{k}-\frac{\eta}{2})}\right)^{N}
=-\prod_{j=1}^{M}\frac{\sinh(u_{k}-u_{j}+\eta)}{\sinh(u_{k}-u_{j}-\eta)} \,, 
\qquad k = 1, \ldots, M \,.
\label{eq:BAXXZ}
\end{equation}
Singular BA solutions appear also for the XXZ spin chain, and the admissibility
of the solution $\{-\frac{\eta}{2},\frac{\eta}{2},u_{1},\dots,u_{M-2}\}$
with pairwise distinct and finite roots can be formulated similarly
to the XXX case as \cite{Nepomechie:2014hma}
\begin{equation}
\frac{\bar{Q}(+\frac{\eta}{2})\bar{Q}(+\frac{3\eta}{2})}
{\bar{Q}(-\frac{\eta}{2})\bar{Q}(-\frac{3\eta}{2})}= (-1)^{N}\,,
\label{physicalXXZ}
\end{equation}
where $\bar{Q}(u)=\prod_{j=1}^{M-2}\sinh(u-u_{j})$. We conjecture 
that the number ${\cal 
N}(N,M)$ of admissible solutions of the BA equations with $M\le N/2$ is given by 
\begin{equation}
	{\cal N}(N,M) = {N \choose M} \,.
	\label{expectedXXZ}
\end{equation}	

In the following we introduce a $Q$-system whose polynomial (in
$t$ and $t^{-1}$) solutions account for the physical solutions.
We work for generic $\eta$, i.e. when $q=e^{\eta}$ is not a root
of unity.

\subsection{$Q$-system and its solution}

For the closed XXZ case, as in the XXX case, $Q_{2,s}=1$ and the
nontrivial $Q$-functions are $Q_{1,s}$ and $Q_{0,s}$; however, there is no
condition on $Q_{1, s \ge M}$.  
We now regard the $Q$-functions
as functions of the argument $t=e^{u}$. Moreover, shifts now denote
$f^{\pm}(t)=f(tq^{\pm\frac{1}{2}})$, and $QQ$-equations are formulated
around each face as 
\begin{equation}
Q_{a+1,s}(t)\, Q_{a,s+1}(t)\propto 
Q_{a+1,s+1}^{+}(t)\, Q_{a,s}^{-}(t)-Q_{a+1,s+1}^{-}(t)\, 
Q_{a,s}^{+}(t) \,.
\label{eq:QQXXZ}
\end{equation}
The initial conditions are 
\begin{equation}
Q_{0,0}(t)=(t-t^{-1})^{N}\,, \qquad 
Q_{1,0}(t)=Q(t)=\prod_{j=1}^{M}(tt_{j}^{-1}-t^{-1}t_{j}) \,.
\label{icQQXXZ}
\end{equation}
Both of these $Q$-functions are polynomial in the variables $t$ and $t^{-1}$. As in
the XXX case, we introduce the analogue of $P$, and then proceed to express all $Q$-functions
in terms of $P$ and $Q$. 

The $QQ$-equation for $(a,s)=(1,0)$ leads again to the discrete
derivative of $Q$:
\begin{equation}
Q_{1,1}(t)=Q_{1,0}^{+}(t)-Q_{1,0}^{-}(t)\equiv Q'_{1,0}(t)=Q'(t) \,.
\end{equation}
However, contrary to the XXX case, the order of $Q_{1,1}$ is the same
as that of $Q$.\footnote{In fact, $Q_{1,n}$ has order $M$, and 
$Q_{0,n}$ has order $N$.}
Since the $QQ$-equations for the XXZ case (\ref{eq:QQXXZ})
are the same as for the XXX case (\ref{eq:QQXXX}), the solutions
are the same, too. In particular, formulas such as (\ref{eq:Q01Q})-(\ref{eq:TfromQ01})
are exactly the same, and polynomiality of $Q_{0,1}$ is equivalent
to the polynomiality of $T$, which gives the BA equations 
(\ref{eq:BAXXZ}). In proceeding as before,
we search for a function $P$ that satisfies 
\begin{equation}
Q_{0,0}=P^{+}Q^{-}-P^{-}Q^{+}\,,
\end{equation}
that is,
\begin{equation}
(t-t^{-1})^{N}=P(tq^{\frac{1}{2}})\, 
Q(tq^{-\frac{1}{2}})-P(tq^{-\frac{1}{2}})\, Q(tq^{\frac{1}{2}}) \,.
\end{equation}
With this $P$ and $Q$, all $Q$-functions can be written as in the XXX 
case (\ref{eq:QQsltnXXX})
\begin{equation}
Q_{1,n}=Q^{(n)}\,, \qquad Q_{0,n} \propto P^{(n)+}Q^{(n)-}-P^{(n)-}Q^{(n)+} 
\,,
\end{equation}
except that the superscript $(n)$ denotes the $n^{th}$ discrete derivative
obtained from multiplicative shifts in $t$, with $f^{\pm}(t)=f(tq^{\pm\frac{1}{2}})$
and $f'(t)=f^{+}(t)-f^{-}(t)$. 

The construction of the function $P$, once written in terms of the
shifts, literally repeats the steps in the XXX case. One first shows
that the only singular solutions are $t=q^{\pm\frac{1}{2}}.$ One
then separates the singular solutions as 
$Q(t)=(t-t^{-1})^{+}(t-t^{-1})^{-}\bar{Q}(t)$,
and performs a partial fraction decomposition of $R(t)$ as 
\begin{equation}
R(t)=\frac{(t-t^{-1})^{N}}{Q^{+}Q^{-}}=\pi+\frac{r_{+}}{\bar{Q}^{+}}+\frac{r_{-}}{\bar{Q}^{-}}
+\frac{a_{+}}{(t-t^{-1})^{++}}+\frac{a_{-}}{(t-t^{-1})^{--}} \,,
\end{equation}
where $\pi(t)$ is polynomial.
From the singularity structure, one can obtain $r_{+}=r^{+}$ and 
$r_{-}=-r^{-}$ for some polynomial function $r(t)$, together with 
\begin{equation}
a_{\pm}=\pm\frac{2^{N}T(q^{\mp\frac{1}{2}})}{(q-q^{-1})^{2}(q^{2}-q^{-2})}
\frac{1}{\bar{Q}(q^{\pm\frac{1}{2}})\bar{Q}(q^{\mp\frac{3}{2}})} \,.
\end{equation}
The important new step now is the discrete integration of 
$(t-t^{-1})^{-1}=\frac{1}{2}(\frac{1}{t-1}+\frac{1}{t+1})$. To this 
end, we define the function $p_{q}(t)$ by
\begin{equation}
p_{q}'(t)=\frac{1}{t-t^{-1}}\,, \qquad 
p_{q}(t)=\frac{1}{2\log q}\left\{ \psi_{q^{-1}}\left(\frac{\log t}{\log q}+\frac{1}{2}\right)
-\psi_{q^{-1}}\left(\frac{\log(-t)}{\log 
q}+\frac{1}{2}\right)\right\} \,,
\label{pqfuncdef}
\end{equation}
where $\psi_{q}(x)$ denotes the $q$-deformed digamma function 
\cite{Thomae:1869, Jackson:1905}, which
satisfies 
\begin{equation}
\psi_{q}(x+1)-\psi_{q}(x)=\frac{\log q}{1-q^{-x}} \,.
\end{equation}
Another step of the XXX case that requires special care when 
generalizing to the XXZ case, which was already addressed in 
\cite{Pronko:1998xa}, is the deformed discrete integration of $\pi$
into $\rho$ such that
\begin{equation}
  \rho '=\pi \,.
\end{equation}
The non-constant terms can be integrated using 
$\left( \frac {t^{n}} {q^{n/2}-q^{-n / 2}}\right)^{\prime}=t^{n}$ for $n\neq 0$. 
However, a constant term cannot be integrated into a polynomial, and
requires instead a function $c_{q}\left( t\right) $ such that
$c_{q}'\left( t\right) =1$. We have 
\begin{equation}
  c_{q}\left( t\right) =\frac {\log t} {\log q} \,,
\end{equation}
up to an additive constant. Hence, the function $P$ finally takes the form
\begin{align}
P=&\rho_0  Q+\alpha\log\left( t\right) Q+(t-t^{-1})^{+}(t-t^{-1})^{-} r 
+ \frac{1}{2}(a_{+}-a_{-})(t-t^{-1})(q^{\frac{1}{2}}-q^{-\frac{1}{2}})\bar{Q}\non \\
&+\frac{1}{2}(a_{+}+a_{-})(p_{q}^{++}+p_{q}^{--}) Q\,,
\label{PXXZclosed}
\end{align}
with $\rho_0$ a polynomial, and $\alpha=\frac {\pi _{0}} {\log q}$ a
constant, where $\pi_0$ is the constant term of $\pi$.  For $N$ odd
$\alpha$ vanishes \cite {Pronko:1998xa}, but it can be non-zero for
$N$ even.  Thus we see that even in absence of strings, $P$ is not
always a polynomial, but what we will call a `quasi-polynomial', i.e. a
polynomial plus $\log t$ times a polynomial.

Quasi-polynomiality of $P$ (in $t,t^{-1}$) requires $a_{+}=-a_{-}$,
which is equivalent to the admissibility of the Bethe roots
(\ref{physicalXXZ}).  As for the polynomiality of the function
$Q_{0,2}$ (and $Q_{0,n}$), one sees that the log part of $P^{\left(
n\right) \pm }$ is $\alpha \log (t)Q^{\left( n\right) \pm }$ so that it
always cancels out in $Q_{0,n}$.  Then one discovers that the
polynomiality of the function $Q_{0,2}$ is also equivalent to the
$a_{+}=-a_{-}$ condition.  Thus, (almost) similarly to the XXX case,
the following statements are equivalent: (i) $P$ is a
quasi-polynomial, (ii) all $Q_{a,s}$-functions are polynomial, (iii)
the roots $\{u_{1},\dots,u_{M}\}$ are admissible solutions of the BA
equations, (iv) the Bethe vector is an eigenvector of the transfer
matrix.

Let us briefly comment on the root of unity 
case $q=e^{i\pi/p}$ with integer $p \ge 2$.  In this case, 
another exact string besides $\pm\eta/2$ becomes possible,
namely, a complete string of length $p$ \cite{Baxter:1972wg, Fabricius:2000yx, 
Baxter:2001sx, Tarasov:2003xz, Gainutdinov:2015vba}, due to
the periodicity of $\sinh$ in the imaginary
direction.  Thus the construction of the function $P$ would involve
more $q$-deformed digamma functions located at the center of these new
exact strings, and the quasi-polynomiality of $P$ would be equivalent to the
cancellation of multiple constants.  It would a priori require more
work to show that their cancellation are equivalent to the $QQ$-relations, 
as we should expect.

\subsubsection{An example}\label{sec:exampleXXZclosed}

We now present an explicit example of 
using the Q-system (\ref{eq:QQXXZ})-(\ref{icQQXXZ}) to compute 
Bethe roots. The 
functions $Q_{0,0}(t)$ and $Q_{1,0}(t)$ are now given by\footnote{In 
order to avoid a vanishing $t^{-M}$ term, we define its coefficient as 
$1/c[0]$, and we consider solutions with finite $c[0]$.}
\begin{verbatim}
Q[0, 0, t_] := (t - t^(-1))^L;  
Q[1, 0, t_] := t^(-M)/c[0] + Sum[c[k] t^(2k - M), {k, 1, M-1}] + t^M;
\end{verbatim}
and the functions $Q_{1,n}(t)$ and $Q_{0,n}(t)$ are given by
\begin{verbatim}
Q[1, n_, t_] := Q[1, n-1, t Exp[eta/2]] - Q[1, n-1, t Exp[-(eta/2)]];
Q[0, n_, t_] := (Q[1, n, t Exp[eta/2]] Q[0, n-1, t Exp[-(eta/2)]] - 
     Q[1, n, t Exp[-(eta/2)]] Q[0, n-1, t Exp[eta/2]])/ Q[1, n-1, t];
\end{verbatim}
Moreover, we define $y_{n}(t)$ by
\begin{verbatim}
y[n_, t_] := PolynomialRemainder[Numerator[t^L Together[Q[0, n, t]]], 
   Denominator[t^L Together[Q[0, n, t]]], t];
\end{verbatim}
where we have inserted $t^L$ factors to ensure that both the numerator
and denominator are polynomials in $t$.
We then solve for the coefficients $c_{k}$ that make $y_{n}(t)$ vanish for 
all values of $t$. From experience, it is enough to consider $n= 1, 
2,\ldots, M$.
\begin{verbatim}
sol = Solve[Table[CoefficientList[y[n, t], t] == 0, {n, 1, M}],  
    Table[c[k], {k, 0, M-1}]]
\end{verbatim}

As before, let us consider the case $N\equiv L=6$ and $M=2$, and 
we now further set $\eta=\log(2)$. The above code 
generates 15 solutions, in agreement with (\ref{expectedXXZ}). We 
can solve for the corresponding Bethe roots similarly as before, for example
\begin{verbatim}
NSolve[(Q[1, 0, Exp[u]] /. sol[[1]]) == 0, u]// Simplify // Flatten
\end{verbatim}
gives $\pm 0.346574$.

\section{Open XXX $Q$-system}\label{sec:openXXX}

We turn now to the open XXX spin chain of length $N$, with Hamiltonian
\be
\HH = \sum_{k=1}^{N-1} \vec \sigma_{k} \cdot \vec \sigma_{k+1} \,,
\label{eq:HopenXXX}
\ee
which is $SU(2)$ invariant. After reviewing its BA solution, we 
propose a corresponding $Q$-system, and argue that all the $Q$'s are polynomial if and 
only if the Bethe state is an eigenstate of the transfer matrix.

\subsection{Review of the algebraic BA solution}

The transfer matrix $\T(u)$ is given by \cite{Sklyanin:1988yz}
\be
\T(u) = \tr_{0} \U_{0}(u) \,, \qquad \U_{0}(u) = \M_{0}(u)\, 
\widehat{\M}_{0}(u) \,,
\label{transferopenXXX}
\ee
where $\M_{0}(u)$ is the monodromy matrix in (\ref{eq:MandT}), and 
$\widehat{\M}_{0}(u)$ is given by
\be
\widehat{\M}(u) = \R_{0 N}(u) \cdots \R_{0 2}(u)\, \R_{0 1}(u) \,.
\ee
The $R$-matrix is again given by (\ref{eq:Rmatrix}). Its boundary 
equivalent, the $K$-matrix, is the identity in the case considered here.
By construction,
the open-chain transfer matrix (\ref{transferopenXXX}) has the 
commutativity property
\be
\left[ \T(u) \,, \T(v) \right] = 0 \,,
\label{commutativityopen}
\ee
and it also has the crossing symmetry
\be
\T(-u) = \T(u) \,.
\label{crosssing}
\ee
The Hamiltonian (\ref{eq:HopenXXX}) is proportional to 
$\frac{d\T(u)}{du}\Big\vert_{u=i/2}$,
up to an additive constant. 

We denote the matrix elements of $\U_{0}(u)$ (\ref{transferopenXXX}) as follows
\be
\U_{0}(u) = \left( \begin{array}{cc}
\A(u) & \B(u) \\
\C(u) & \frac{u^{-}}{u}  \D(u) + \frac{i}{2u} \A(u) 
\end{array} \right) \,.
\ee 
The reference state (\ref{reference})
is annihilated by $\C(u)$, and is an eigenstate of $\A(u)$ and 
$\D(u)$, with
\be
\A(u)|0\rangle  = (u^{+})^{2N} |0\rangle\,, \qquad \D(u)|0\rangle  = 
(u^{-})^{2N} |0\rangle \,.
\ee
The Bethe states are defined by
\be
|u_{1} \ldots u_{M} \rangle = \prod_{k=1}^{M} \B(u_{k})  |0\rangle \,.
\label{BethestateopenXXX} 
\ee
The Bethe states satisfy (for any $\{ u_{1}\,, \ldots \,, u_{M} \}$) the off-shell 
relation
\be
\T(u) |u_{1} \ldots u_{M} \rangle = T(u) 
|u_{1} \ldots u_{M} \rangle + 
\sum_{j=1}^{M} F_{j}|u, u_{1} \ldots \hat u_{j} 
\ldots u_{M} \rangle \,,
\label{offshellXXX}
\ee
where $\hat u_{j}$ is omitted. Moreover,
$T(u)$ is given by the $TQ$-relation
\be
u\, T(u)\, Q(u) = 
(u^{+})^{2N+1}\, Q^{--}(u) +
 (u^{-})^{2N+1}\, Q^{++}(u) \,, \quad 
Q(u) =\prod_{k=1}^{M}\left(u-u_{k}\right)\left(u+u_{k}\right) \,,
\label{TQopenXXX}
\ee
and the coefficients $F_{j}$ of the ``unwanted'' terms are given by 
\begin{align}
F_{j} &= \frac{2i u^{+} (u_{j}-\frac{i}{2})}{u_{j} (u-u_{j}) (u+u_{j})}
\Bigg[(u_{j}+\tfrac{i}{2})^{2N}
\prod_{\scriptstyle{k \ne j}\atop \scriptstyle{k=1}}^M
\frac{(u_{j}-u_{k}-i) (u_{j}+u_{k}-i)}
{(u_{j}-u_{k})(u_{j}+u_{k})} \non \\
&\qquad - (u_{j}-\tfrac{i}{2})^{2N}
\prod_{\scriptstyle{k \ne j}\atop \scriptstyle{k=1}}^M
\frac{(u_{j}-u_{k}+i) (u_{j}+u_{k}+i)}
{(u_{j}-u_{k})(u_{j}+u_{k})} 
\Bigg] \,.
\label{FjXXX}
\end{align}
We again write $f^\pm(u) = f(u \pm \frac{i}{2})$, as in the closed XXX case.
Note that both $Q$ and $T$ are even functions of $u$
\be
Q(-u) = Q(u)\,, \qquad T(-u) = T(u) \,.
\ee 
Substituting $u=u_{j}$ in the $TQ$-relation (\ref{TQopenXXX}), we see 
that the LHS vanishes, and we obtain
\be
(u_{j}+\tfrac{i}{2})^{2N+1}\, \prod_{k=1}^{M} 
(u_{j}-u_{k}-i)(u_{j}+u_{k}-i) +
(u_{j}-\tfrac{i}{2})^{2N+1}\, \prod_{k=1}^{M} 
(u_{j}-u_{k}+i)(u_{j}+u_{k}+i) = 0 \,, \quad j = 1, \ldots, M \,.
\label{BAEopenXXX1}
\ee 
If $u_{j} \ne \pm \frac{i}{2}$, then these equations are equivalent to
the BA equations
\be
\left(\frac{u_{j}+\tfrac{i}{2}}{u_{j}-\tfrac{i}{2}} \right)^{2N}
= \prod_{\scriptstyle{k \ne j}\atop \scriptstyle{k=1}}^M
\frac{(u_{j}-u_{k}+i)(u_{j}+u_{k}+i)}{(u_{j}-u_{k}-i)(u_{j}+u_{k}-i)}\,, \quad j = 1, \ldots, M \,.
\label{BAEopenXXX2}
\ee
The BA equations have the reflection symmetry $u_{j} \mapsto
-u_{j}$, while keeping the other $u$'s (i.e. 
$u_{k}$ with $k \ne j$) unchanged. Hence, without loss of generality, 
we henceforth assume that $\Re e(u_{j} ) > 0$, or $\Re e(u_{j} ) = 0$ and 
$\Im m(u_{j} ) \ge 0$.

Roughly speaking, if $\{ u_{1}\,, \ldots \,, u_{M} \}$ satisfy the 
BA equations (\ref{BAEopenXXX2}), then all $F_{j}=0$; 
i.e., the ``unwanted'' terms in the off-shell 
relation (\ref{offshellXXX}) vanish, hence the Bethe 
state (\ref{BethestateopenXXX}) is an eigenstate of the transfer 
matrix, with corresponding eigenvalue $T(u)$. 
However, there are some important caveats.
We argue in Appendix \ref{sec:exceptional} that certain 
``exceptional'' solutions of the BA 
equations (namely $0$ and $\pm \frac{i}{2}$) do {\em not} lead to 
eigenstates of the transfer matrix. Moreover, we make the 
standard assumption (supported by numerical evidence, see e.g. 
\cite{Gainutdinov:2015vba}) that the Bethe roots are 
pairwise distinct, i.e. $u_{j} \ne u_{k}$ if $j\ne k$.

We therefore define an {\em admissible} 
solution $\{ u_{1}\,, \ldots \,, u_{M} \}$ of the BA 
equations (\ref{BAEopenXXX2}), such that all the $u_{j}$'s are 
finite, not equal to $\pm \frac{i}{2}$ or 0,
and pairwise distinct (no two are equal), and each $u_{j}$
satisfies either
\be
\Re e(u_{j} ) > 0   
\label{admissibleXXXa}
\ee
or
\be
\Re e(u_{j} ) = 0  \qquad \mbox{ and  } \qquad \Im m(u_{j} ) >
0 \,. 
\label{admissibleXXXb}
\ee
The set $\{ u_{1}\,, \ldots \,, u_{M}\}$ is an admissible solution of 
the BA equations if and only if the Bethe state $|u_{1} \ldots 
u_{M} \rangle$ is an eigenstate of the transfer matrix $\T(u)$. We 
emphasize that, for the open XXX chain, there are no physical singular solutions of the 
Bethe equations -- all the singular solutions are unphysical. 
The number ${\cal N}(N,M)$ of admissible solutions of the BA 
equations with $M\le N/2$  has been conjectured \cite{Gainutdinov:2015vba} to be given by 
\begin{equation}
	{\cal N}(N,M) = {N \choose M} - {N \choose M-1} \,.
	\label{expectedXXXopen}
\end{equation}	

\subsection{$Q$-system}

We propose the following $Q$-system 
\be
u\, Q_{a+1,s}(u)\, Q_{a,s+1}(u)\propto Q_{a+1,s+1}^{+}(u)\, Q_{a,s}^{-}(u)
- Q_{a+1,s+1}^{-}(u)\, Q_{a,s}^{+}(u)\,,
\label{QQopenXXX}
\ee
where the nontrivial $Q$-functions for given values of $N$ and $M$ are again 
defined on the Young tableau in Fig. \ref{QQdiagramXXX},
with the boundary conditions $Q_{2,s}=1\,,\  Q_{1,s\ge M} = 1$,  
and with the initial condition
\be
Q_{0,0}(u) = u^{2N}  \qquad  \mbox{ with  } \qquad 
Q_{1,0}(u) = Q(u) = 
\prod_{k=1}^{M}\left(u-u_{k}\right)\left(u+u_{k}\right) \,.
\label{icQQopenXXX}
\ee 
In contrast to the $Q$-system for periodic XXX (\ref{eq:QQXXX}), there is an extra 
factor of $u$ on the LHS of (\ref{QQopenXXX}), and the $Q$'s are {\em 
even} functions of $u$. The degree of the 
polynomial $Q_{a,s}(u)$ is \emph{doubled} with respect to the 
periodic XXX case (namely, \emph{twice} the number of boxes in the Young tableau
to the right and top of the vertex $(a,s)$). \footnote{Equivalently, 
the Q-functions are polynomials in the variable $u^2$ of the same degree as in the periodic XXX case.}
We claim that all the $Q$'s are polynomial if and 
only if the Bethe state $|u_{1} \ldots u_{M} \rangle$ (\ref{BethestateopenXXX})
is an eigenstate of the transfer matrix $\T(u)$ (\ref{transferopenXXX}).

Before entering into the proof, let us quickly check that this $Q$-system 
indeed leads to the correct BA 
equations for $\{ u_{1}\,, \ldots \,, u_{M}\}$. We write the $QQ$-equations for $(a,s) = (0,0)$:
\be
u\, Q_{1,0}\, Q_{0,1} \propto Q^{+}_{1,1}\, 
Q^{-}_{0,0} - Q^{-}_{1,1}\, Q^{+}_{0,0} \,,
\label{a0s0}
\ee
and for $(a,s) = (1,0)$:
\be
u\, Q_{2,0}\, Q_{1,1} \propto Q^{+}_{2,1}\, 
Q^{-}_{1,0} - Q^{-}_{2,1}\, Q^{+}_{1,0} \,.
\ee
Since $Q_{2,0} = Q_{2,1}  =  1$, the latter reduces to 
\be
u\, Q_{1,1} \propto 
Q^{-} - Q^{+} \,.
\label{a1s0}
\ee
Performing the shifts $u \mapsto u \pm \frac{i}{2}$ in (\ref{a1s0}) and evaluating at 
$u=u_{j}$, we obtain
\be
(u_{j} +\tfrac{i}{2})\,  Q_{1,1}^{+}(u_{j}) \propto 
- Q^{++}(u_{j}) \,, \qquad 
(u_{j} -\tfrac{i}{2})\,  Q_{1,1}^{-}(u_{j}) \propto 
Q^{--}(u_{j})  \,,
\label{step}
\ee
since $Q(u_{j}) = 0$. Moreover,  evaluating (\ref{a0s0}) at $u=u_{j}$ gives 
\be
Q^{+}_{1,1}(u_{j})\, 
Q^{-}_{0,0}(u_{j}) = Q^{-}_{1,1}(u_{j})\, Q^{+}_{0,0}(u_{j}) \,,
\ee
Substituting (\ref{step}) into the above relation gives
\be
-(u_{j} -\tfrac{i}{2})\, Q^{++}(u_{j})\, Q^{-}_{0,0}(u_{j})  
= (u_{j} +\tfrac{i}{2})\, Q^{--}(u_{j})\, Q^{+}_{0,0}(u_{j}) \,,
\label{BEprelim}
\ee
which coincides with the BA equations (\ref{BAEopenXXX1}).

\subsubsection{$Q_{a,s}$ in terms of $Q$ and $P$}

We now solve the $Q$-system (\ref{QQopenXXX}) in terms of $Q(u)$ and a function $P(u)$, such that 
polynomiality of $P(u)$ implies polynomiality of all the $Q$'s. We 
define $P(u)$ by \footnote{As in the closed-chain case, for given $Q$, this equation 
does not uniquely define $P$: if $P(u)$ is a solution of (\ref{PopenXXX}), then so is 
$P(u) + \alpha\, Q(u)$, for any constant value of $\alpha$.}
\be
P^{+} Q^{-} - P^{-} Q^{+} = u\, Q_{0,0} \,,
\label{PopenXXX}
\ee
where $Q_{0,0}$ is given by (\ref{icQQopenXXX}). It follows from 
(\ref{a0s0}) and (\ref{a1s0}) that
\be
Q_{1,1} \propto \frac{Q'}{u} = DQ \,, \qquad 
u\, Q_{0,1} \propto (DP)^{+} (DQ)^{-} - (DP)^{-} (DQ)^{+} \,,
\label{Q1openXXX}
\ee 
where we have used the following compact notation for discrete derivatives 
with certain $1/u$ factors
\begin{align}
Df &= \frac{1}{u}(f^{+} - f^{-}) = \frac{f'}{u} \,, \non \\
D^{2}f &= \frac{1}{u}\left[(Df)^{+} - (Df)^{-} \right] \,, \ldots \non \\
D^{n}f &= \frac{1}{u}\left[(D^{n-1}f)^{+} - (D^{n-1}f)^{-} \right] \,.
\label{DXXX}
\end{align}
Similarly, we obtain
\be
Q_{1,2} \propto D^{2}Q \,, \qquad 
u\, Q_{0,2} \propto (D^{2}P)^{+} (D^{2}Q)^{-} - (D^{2}P)^{-} 
(D^{2}Q)^{+} \,,
\label{Q2openXXX}
\ee 
and in general
\be
Q_{1,n} \propto D^{n}Q \,, \qquad 
u\, Q_{0,n} \propto (D^{n}P)^{+} (D^{n}Q)^{-} - (D^{n}P)^{-} 
(D^{n}Q)^{+} \,.
\label{QnopenXXX}
\ee 

Since both $Q(u)$ and $P(u)$ are even functions of $u$, it follows 
that $DQ$ and $DP$ are also even functions of $u$. Hence, if $P(u)$ 
is a polynomial function of $u$, then the RHS of the second equation 
in (\ref{Q1openXXX}) is 
divisible by $u$, thus $Q_{0,1}$ is polynomial; and, from 
(\ref{QnopenXXX}), we similarly conclude that all the $Q$'s are polynomial.

We observe, similarly to the closed-chain case, that the $TQ$-equation 
(\ref{TQopenXXX}) together with the definition of $P$ (\ref{PopenXXX}) imply
\be
u\, T = P^{++} Q^{--} - P^{--} Q^{++} \,.
\label{TPQopenXXX}
\ee
It follows that $P$ is also a solution of the $TQ$-equation
\be
u\, T\, P = (u^{+})^{2N+1}\, P^{--} + (u^{-})^{2N+1}\, P^{++} \,.
\label{TPopenXXX}
\ee
Hence, (\ref{PopenXXX}) can be regarded as the Wronskian relation 
obeyed by the two solutions $Q$ and $P$ of the $TQ$-equation (\ref{TQopenXXX}).

\subsubsection{Construction of $P$}

We now construct the function $P(u)$ for a set $\{ u_{1}\,, \ldots \,, u_{M}\}$, 
and argue that $P(u)$ is polynomial if and only if $\{ u_{1}\,, \ldots \,, 
u_{M}\}$ is an admissible solution of the BA equations.

The construction of the $P$-function for the open chain is similar to that
for the closed chain, but with some significant differences. In the 
presence of one singular root $\tfrac{i}{2}$ and one zero root 0, 
the $Q$-function takes the form \footnote{The cases of either 
one singular root or one zero root are essentially special cases, for 
which $b_{\pm} = c_{\pm}=0$ or $a_{\pm} = 0$ in (\ref{RpfopenXXX}) below, 
respectively.}
\be
Q(u) = u^{+} u^{-} u^{2}\bar{Q}(u)\,, \qquad \bar{Q}(u) =
\prod_{\scriptstyle{u_{k} \ne 
\frac{i}{2}, 0}\atop \scriptstyle{k=1}}^M (u-u_{k})(u+u_{k}) \,.
\ee 
We define the function $R(u)$
\be
R = \frac{u^{2N+1}}{Q^{+} Q^{-}} = \frac{u^{2N-1}}{u^{++} u^{--} 
(u^{+} u^{-})^{2}
\bar{Q}^{+} \bar{Q}^{-}} \,,
\label{RopenXXX}
\ee
which is related to $P(u)$ defined in (\ref{PopenXXX}) by
\be
R = \left(\frac{P}{Q}\right)' = \frac{P^{+}}{Q^{+}} - 
\frac{P^{-}}{Q^{-}} \,.
\label{RPQopenXXX}
\ee
Decomposing (\ref{RopenXXX}) in partial fractions, we obtain
\be
R = \pi + \frac{q_{+}}{\bar{Q}^{+}} + \frac{q_{-}}{\bar{Q}^{-}} 
+ \frac{a_{+}}{u^{++}} + \frac{b_{+}}{u^{+}} + \frac{c_{+}}{(u^{+})^{2}} 
+ \frac{a_{-}}{u^{--}} + \frac{b_{-}}{u^{-}} + \frac{c_{-}}{(u^{-})^{2}} \,,
\label{RpfopenXXX}
\ee
where $\pi$ is a polynomial of order $2N-4M+1$, $q_{\pm}$ are 
polynomials of degree less than that of $\bar{Q}$, and $a_{\pm}\,, 
b_{\pm}\,, c_{\pm}$ are constants. Note that $a_{\pm}$ arise from 
the presence of the singular root, while $b_{\pm}\,, c_{\pm}$ are due 
to the presence of the zero root.
From the $TQ$-relation (\ref{TQopenXXX}), we obtain
\be
R^{+} + R^{-} = \frac{u\, T}{Q^{++} Q^{--}} = 
\frac{u\, T}{u^{+} u^{-} u^{+++} u^{---} (u^{++} u^{--})^{2}\bar{Q}^{++} \bar{Q}^{--}} \,.
\label{RpRmopenXXX}
\ee
We now evaluate the LHS of (\ref{RpRmopenXXX}) using (\ref{RpfopenXXX}), 
and consider the values of $u$ where singularities could arise.
The RHS of (\ref{RpRmopenXXX}) has no singularities at the 
zeros of $\bar{Q}$ (recall that $T(u)$ is regular for values of 
$u$ corresponding to admissible Bethe roots), hence
\be
q_{+} = q^{+} \,, \qquad q_{-} = -q^{-} \,,
\ee
for some polynomial $q(u)$. From the residues of (\ref{RpRmopenXXX})  
at $u=\mp \frac{i}{2}$, we obtain
\be
a_{\pm} = \frac{4 T(\mp\frac{i}{2})}{9\bar{Q}(\pm\frac{i}{2}) 
\bar{Q}(\mp\frac{3i}{2})} = \frac{8 (-1)^{N+1}}{9\bar{Q}(\pm\frac{i}{2}) 
\bar{Q}(\mp\frac{3i}{2})}\,.
\ee
Since $Q(u)$ is an even function of $u$, we conclude 
-- in significant contrast from the closed-chain case -- that $a_{+} = 
a_{-} \equiv a$. From the residues of (\ref{RpRmopenXXX})  
at $u=\mp i$, we can obtain expressions for $c_{\pm}$ and $b_{\pm}$,
and we find that
\be
c \equiv c_{+} = -c_{-}\,, \qquad b \equiv b_{+} = b_{-}\,,
\ee 
which is consistent with the constraints coming from the residues 
of (\ref{RpRmopenXXX}) at $u=0$ 
(note that the presence of a zero root implies that $T(u)$ has a double 
pole at $u=0$).

We write the polynomial $\pi$ in (\ref{RpfopenXXX}) as
\be
\pi = \rho' = \rho^{+} - \rho^{-} \,,
\ee
where $\rho$ is a polynomial. Recalling the definition of the 
function $p(u)$ (\ref{pfuncdef}), we see that
\be
R = \left( \rho + \frac{q}{\bar{Q}} + a (p^{++} + p^{--})+ b (p^{+} + p^{-}) + 
\frac{c}{u^{2}} \right)'  \,.
\label{RnewopenXXX}
\ee 
It immediately follows from (\ref{RPQopenXXX}) that the $P$-function is given by
\be
P = \rho\, Q + u^{+} u^{-} u^{2} q + c\, u^{+} u^{-} \bar{Q}
+ a (p^{++} + p^{--}) Q + b (p^{+} + p^{-}) Q\,,
\ee
which is a polynomial iff $a=b=0$. 
That is, the $P$-function is 
polynomial iff there is no singular root and no zero root, in which case $\{ u_{1}\,, 
\ldots \,, u_{M}\}$ is admissible. Moreover, $Q_{0,2}$ is polynomial 
iff $a=b=0$. 

The proof of the $Q$-system (\ref{QQopenXXX}) is now complete, since we have argued 
as before that the following statements are equivalent: (i) $P$ is a polynomial, (ii) all 
$Q_{a,s}$-functions are polynomial, (iii) the roots $\{u_{1},\dots,u_{M}\}$ are
admissible solutions of the BA equations, (iv) the Bethe vector is an
eigenvector of the transfer matrix. 

\subsubsection{An example}\label{sec:exampleXXXopen}

We now present an explicit example of 
using the Q-system (\ref{QQopenXXX})-(\ref{icQQopenXXX}) to compute 
Bethe roots. The 
code is similar to the one in Sec. \ref{sec:exampleXXXclosed}. The 
functions $Q_{0,0}(u)$ and $Q_{1,0}(u)$ are now given by
\begin{verbatim}
Q[0, 0, u_] := u^(2L);  
Q[1, 0, u_] := Sum[c[k] u^(2k), {k, 0, M-1}] + u^(2M);
\end{verbatim}
while $Q_{1,n}(u)$ and $Q_{0,n}(u)$ are given by
\begin{verbatim}
Q[1, n_, u_] := (Q[1, n-1, u + I/2] - Q[1, n-1, u - I/2])/u; 
Q[0, n_, u_] := (Q[1, n, u + I/2] Q[0, n-1, u - I/2] - 
   Q[1, n, u - I/2] Q[0, n-1, u + I/2])/( u Q[1, n-1, u]);
\end{verbatim}
We define $y_{n}(u)$ and solve for the coefficients $c_{k}$ exactly as in Sec. 
\ref{sec:exampleXXXclosed}.

For the case $N\equiv L=6, M=2$, the above code 
generates 9 solutions, in agreement with (\ref{expectedXXXopen}). 
For example, for the first solution, the corresponding Bethe roots are 
given by $0.301932, 1.26627$.

\section{Open quantum-group-invariant XXZ $Q$-system}\label{sec:openXXZ}

We now generalize the preceding results to the open 
quantum-group-invariant XXZ spin chain, whose Hamiltonian is 
given by \cite{Pasquier:1989kd}
\be
\HH = \sum_{k=1}^{N-1} \left[ \sigma^x_k \sigma^x_{k+1} +
\sigma^y_k \sigma^y_{k+1} + {1\over 2}( q + q^{-1}) \sigma^z_k
\sigma^z_{k+1}
\right]  - {1\over 2}( q - q^{-1})
\bigl( \sigma^z_1 - \sigma^z_N \bigr) \,.
\label{eq:HopenXXZ}
\ee
For simplicity, we restrict to generic values of $q=e^{\eta}$.

\subsection{Review of the algebraic BA solution}

The transfer matrix is now given by \cite{Sklyanin:1988yz}
\be
\T(u) =  \tr_{0} \K^{L}_{0}(u)\, \U_{0}(u) \,, \qquad 
\U_{0}(u) =  \M_{0}(u)\, \K^{R}_{0}(u)\, \widehat{\M}_{0}(u) 
\,, 
\label{transferXXZ}
\ee
where the $R$-matrix is again given by (\ref{eq:RXXZ}),
and the left and right $K$-matrices (solutions of boundary 
Yang-Baxter equations) are given by the diagonal matrices
\be
\K^{L}(u)=\diag(e^{-u-\frac{\eta}{2}}\,, e^{u+\frac{\eta}{2}}) \,, \qquad 
\K^{R}(u)=\diag(e^{u-\frac{\eta}{2}}\,, e^{-u+\frac{\eta}{2}})\,.
\ee 
The transfer matrix (\ref{transferXXZ}) has the commutativity property 
(\ref{commutativityopen}) as well as the crossing symmetry (\ref{crosssing}).
The Hamiltonian (\ref{eq:HopenXXZ}) is proportional to 
$d\T(u)/du\Big\vert_{u=\eta/2}$, 
up to an additive constant.

We define the elements of $\U_{0}(u)$ (\ref{transferXXZ}) as 
follows
\be
\U_{0}(u) = \left( \begin{array}{cc}
e^{u-\frac{\eta}{2}} \A(u) & \B(u) \\
\C(u) & \frac{e^{-u-\frac{\eta}{2}}\sinh(2u-\eta)}{\sinh(2u)}  \D(u) 
+ \frac{e^{u-\frac{\eta}{2}}\sinh(\eta)}{\sinh(2u)}  \A(u) 
\end{array} \right) \,.
\ee 
The reference state (\ref{reference})
is annihilated by $\C(u)$, and is an eigenstate of $\A(u)$ and $\D(u)$
\be
\A(u)|0\rangle  = \sinh^{2N}(u+\tfrac{\eta}{2}) |0\rangle\,, \qquad 
\D(u)|0\rangle  = \sinh^{2N}(u-\tfrac{\eta}{2}) |0\rangle \,.
\ee
The Bethe states are again defined by
\be
|u_{1} \ldots u_{M} \rangle = \prod_{k=1}^{M} \B(u_{k})  |0\rangle \,.
\label{BethestateopenXXZ} 
\ee
The off-shell equation is
\be
\T(u) |u_{1} \ldots u_{M} \rangle = T(u) 
|u_{1} \ldots u_{M} \rangle + 
\sum_{j=1}^{M} F_{j}|u, u_{1} \ldots \hat u_{j} 
\ldots u_{M} \rangle \,,
\label{offshellXXZ}
\ee
where $T(u)$ is given by the $TQ$-relation
\be
\sinh(2u)\, T(u)\, Q(u) = 
\sinh(2u+\eta)\, \sinh^{2N}(u+\tfrac{\eta}{2})\, Q(u-\eta) +
\sinh(2u-\eta)\, \sinh^{2N}(u-\tfrac{\eta}{2})\, Q(u+\eta)\,, 
\label{TQopenXXZ}
\ee
with
\be
Q(u) =\prod_{k=1}^{M}\sinh(u-u_{k})\, \sinh(u+u_{k}) \,,
\ee
and $F_{j}$ is given by
\begin{align}
F_{j} &= \frac{\sinh(2 u+\eta) \sinh(2 u_{j}-\eta) \sinh(\eta)}
{\sinh(2 u_{j}) \sinh(u-u_{j}) \sinh(u+u_{j}) }
\Bigg[\sinh^{2N}(u_{j}+\tfrac{\eta}{2})
\prod_{\scriptstyle{k \ne j}\atop \scriptstyle{k=1}}^M
\frac{\sinh(u_{j}-u_{k}-\eta) \sinh(u_{j}+u_{k}-\eta)}
{\sinh(u_{j}-u_{k}) \sinh(u_{j}+u_{k})} \non \\
&\qquad - \sinh^{2N}(u_{j}-\tfrac{\eta}{2})
\prod_{\scriptstyle{k \ne j}\atop \scriptstyle{k=1}}^M
\frac{\sinh(u_{j}-u_{k}+\eta) \sinh(u_{j}+u_{k}+\eta)}
{\sinh(u_{j}-u_{k}) \sinh(u_{j}+u_{k})} 
\Bigg] \,.
\label{FjXXZ}
\end{align}
Substituting $u=u_{j}$ in the $TQ$-equation (\ref{TQopenXXZ}), we see 
that the LHS vanishes, and we obtain
\begin{align}
&\sinh(2 u_{j} + \eta)\, \sinh^{2N}(u_{j}+\tfrac{\eta}{2})\, \prod_{k=1}^{M} 
\sinh(u_{j}-u_{k}-\eta)\, \sinh(u_{j}+u_{k}-\eta) \label{BAEopenXXZ1} \\
& + \sinh(2 u_{j} - \eta)\,\sinh^{2N}(u_{j}-\tfrac{\eta}{2})\, \prod_{k=1}^{M} 
\sinh(u_{j}-u_{k}+\eta) \sinh(u_{j}+u_{k}+\eta) = 0 \,, \quad j = 1, \ldots, M 
\,. \non
\end{align}
If $u_{j} \ne \pm \frac{\eta}{2}$, then these equations are equivalent to
the BA equations
\be
\left(\frac{\sinh(u_{j}+\tfrac{\eta}{2})}{\sinh(u_{j}-\tfrac{\eta}{2})} \right)^{2N}
= \prod_{\scriptstyle{k \ne j}\atop \scriptstyle{k=1}}^M
\frac{\sinh(u_{j}-u_{k}+\eta)\sinh(u_{j}+u_{k}+\eta)}
{\sinh(u_{j}-u_{k}-\eta)\sinh(u_{j}+u_{k}-\eta)}\,, \quad j = 1, \ldots, M \,.
\label{BAEopenXXZ2}
\ee
The BA equations have the reflection symmetry $u_{j} \mapsto
-u_{j}$ (while keeping the other $u$'s unchanged), as well as the 
periodicity $u_{j} \mapsto u_{j} + i \pi$. 

We must exclude solutions with roots $0$, $\pm \frac{i \pi}{2}$ and 
$\pm \frac{\eta}{2}$, see Sec. \ref{subsec:exceptionalXXZ}.
We therefore define an {\em admissible} 
solution $\{ u_{1}\,, \ldots \,, u_{M} \}$ of the BA 
equations (\ref{BAEopenXXZ2}), such that all the $u_{j}$'s are 
finite, not equal to $0$, $\pm \frac{i \pi}{2}$ or $\pm \frac{\eta}{2}$,
pairwise distinct, and each $u_{j}$
satisfies either
\be
\Re e(u_{j} ) > 0   \qquad \mbox{ and  } \qquad 
-\frac{\pi}{2} < \Im m(u_{j} ) \le \frac{\pi}{2}
\label{admissibleXXZa}
\ee
or
\be
\Re e(u_{j} ) = 0  \qquad \mbox{ and  } \qquad 0 < \Im m(u_{j} ) < 
\frac{\pi}{2}\,,   
\label{admissibleXXZb}
\ee
The set $\{ u_{1}\,, \ldots \,, u_{M}\}$ is an admissible solution of 
the BA equations if and only if the Bethe state $|u_{1} \ldots 
u_{M} \rangle$ is an eigenstate of the transfer matrix $\T(u)$. 
The number ${\cal N}(N,M)$ of admissible solutions of the BA 
equations with $M\le N/2$ has been conjectured \cite{Gainutdinov:2015vba} to be given 
again by (\ref{expectedXXXopen}).

\subsection{$Q$-system}

We propose the following $Q$-system\footnote{As for the periodic XXZ 
case, here again $t=e^{u}$, and $f^{\pm}(t)=f(tq^{\pm\frac{1}{2}})$.}
\be
(t^{2} - t^{-2})\, Q_{a+1,s}(t)\, Q_{a,s+1}(t)\propto 
Q_{a+1,s+1}^{+}(t)\, Q_{a,s}^{-}(t)-Q_{a+1,s+1}^{-}(t)\, 
Q_{a,s}^{+}(t) \,,
\label{QQopenXXZ}
\ee
where the nontrivial $Q$-functions for given values of $N$ and $M$ are again 
defined on the Young tableau in Fig. \ref{QQdiagramXXX},
with the boundary conditions $Q_{2,s}=1\,,\  Q_{1,s \ge M} = 1$,  
and with the initial condition
\be
Q_{0,0}(t) = (t - t^{-1})^{2N}  \qquad  \mbox{ with  } \qquad 
Q_{1,0}(t) = Q(t) = 
\prod_{k=1}^{M}\left(t t_{k}^{-1} - t^{-1} t_{k} \right)
\left(t t_{k} - t^{-1} t_{k}^{-1} \right) \,.
\label{icQQopenXXZ}
\ee 
In contrast to the periodic XXZ case (\ref{eq:QQXXZ}), there is an 
extra factor $(t^{2} - t^{-2}) \propto \sinh(2u)$ on the LHS of 
(\ref{QQopenXXZ}). This $Q$-system indeed leads to the BA equations 
(\ref{BAEopenXXZ1}), as can be seen by following the same logic (\ref{a0s0}) 
- (\ref{BEprelim}) of the rational case. The degree of the 
polynomials (highest power of $t\equiv e^u$) is the same as in the open XXX case.

\subsubsection{$Q_{a,s}$ in terms of $Q$ and $P$}

We now define $P$ by
\be
P^{+} Q^{-} - P^{-} Q^{+} = (t^{2} - t^{-2})\, Q_{0,0} \,,
\label{PopenXXZ}
\ee
where $Q_{0,0}$ is given by (\ref{icQQopenXXZ}). Similarly to the 
rational case (\ref{QnopenXXX}), we find that the $Q$-system 
(\ref{QQopenXXZ}) implies that
\be
Q_{1,n} \propto D^{n}Q \,, \qquad 
(t^{2} - t^{-2})\, Q_{0,n} \propto (D^{n}P)^{+} (D^{n}Q)^{-} - (D^{n}P)^{-} 
(D^{n}Q)^{+} \,,
\label{QnopenXXZ}
\ee 
where $D$ is now defined by
\begin{align}
Df &= \frac{1}{(t^{2} - t^{-2})}(f^{+} - f^{-}) = \frac{f'}{(t^{2} - t^{-2})} \,, \non \\
D^{2}f &= \frac{1}{(t^{2} - t^{-2})}\left[(Df)^{+} - (Df)^{-} \right] \,, \ldots \non \\
D^{n}f &= \frac{1}{(t^{2} - t^{-2})}\left[(D^{n-1}f)^{+} - 
(D^{n-1}f)^{-} \right] \,,
\end{align}
cf. (\ref{DXXX}).
Note that $P$ is also a solution of the $TQ$-relation 
(\ref{TQopenXXZ})
\be
\sinh(2u)\, T\, P = \sinh(2u+\eta)\, Q_{0,0}^{+}\, P^{--} + 
\sinh(2u-\eta)\, Q_{0,0}^{-}\, P^{++} \,,
\ee
cf. (\ref{TPopenXXX}).

\subsubsection{Construction of $P$}

The construction of $P$ parallels the rational case. In the presence 
of one singular root ($u=\frac{\eta}{2}$, $t = q^{\frac{1}{2}}$) and 
one zero root ($u=0$, $t=1$; the case $u=\frac{i \pi}{2}$, $t=-1$ is similar), the 
$Q$-function becomes
\be
Q(t) = (t - t^{-1})^{+} (t - t^{-1})^{-} (t - t^{-1})^{2}\, 
\bar{Q}(t)\,.
\ee
We now define $R(t)$ as
\be
R(t) = \frac{(t^{2} - t^{-2}) (t - t^{-1})^{2N}}{Q^{+}\, Q^{-}} 
= \frac{(t^{2} - t^{-2}) (t - t^{-1})^{2N-2}}
{(t - t^{-1})^{++} (t - t^{-1})^{--} [(t - t^{-1})^{+} (t - 
t^{-1})^{-} ]^{2}\bar{Q}^{+}\, \bar{Q}^{-}} \,,
\ee
which is related to $P$ (\ref{PopenXXZ}) by $R=(\frac{P}{Q})'$.
We decompose $R$ as follows
\begin{align}
R &= \pi + \frac{r_{+}}{\bar{Q}^{+}} + \frac{r_{-}}{\bar{Q}^{-}} 
+ \frac{a_{+}}{(t - t^{-1})^{++}} + \frac{b_{+}}{(t - t^{-1})^{+}} + 
\frac{c_{+}}{[(t - t^{-1})^{+}]^{2}} \non \\
& + \frac{a_{-}}{(t - t^{-1})^{--}} + \frac{b_{-}}{(t - t^{-1})^{-}} + 
\frac{c_{-}}{[(t - t^{-1})^{-}]^{2}} \,,
\label{RpfopenXXZ}
\end{align}
where $\pi$ is polynomial. From the $TQ$-relation (\ref{TQopenXXZ}), 
we obtain 
\begin{align}
R^{+} + R^{-} &= \frac{2^{2N}(t^{2} - t^{-2}) \, T}{Q^{++} Q^{--}} \non\\
&= \frac{2^{2N}(t^{2} - t^{-2}) \, T}{(t - t^{-1})^{+} (t - t^{-1})^{-} (t - t^{-1})^{+++} (t - t^{-1})^{---} 
[(t - t^{-1})^{++} (t - t^{-1})^{--}]^{2}\bar{Q}^{++} \bar{Q}^{--}} \,.
\label{RpRmopenXXZ}
\end{align}
From the singularity structure of this equation and the fact that 
$Q(t)$ and $T(t)$ are invariant under $t \mapsto t^{-1}$, we again obtain 
\be
r_{+} = r^{+} \,, \qquad r_{-} = -r^{-} \,,
\ee
where $r(t)$ is a polynomial in $t$ and $t^{-1}$, and 
\be
a_{+} = a_{-} \equiv a\,, \qquad c \equiv c_{+} = -c_{-}\,, \qquad b 
\equiv b_{+} = b_{-}\,.
\ee
The expression (\ref{RpfopenXXZ}) for $R$ can therefore be rewritten in the form
\be
R = \left( \rho + \frac{r}{\bar{Q}} + a (p_{q}^{++} + p_{q}^{--})+ b 
(p_{q}^{+} + p_{q}^{-}) + \frac{c}{(t-t^{-1})^{2}} \right)'  \,,
\ee 
where $p_{q}(t)$ is defined in (\ref{pqfuncdef}), cf. 
(\ref{RnewopenXXX}), and $\rho'=\pi$.\footnote{In contrast with the closed XXZ 
case, a term in $P$ of the form $\alpha\log\left( t\right) Q$ as in (\ref{PXXZclosed}) 
is absent.} 
Since $R=(\frac{P}{Q})'$, we conclude that $P$ is 
given by
\begin{align}
P &= \rho\, Q + (t-t^{-1})^{+} (t-t^{-1})^{-} (t-t^{-1})^{2} r 
+ c\, (t-t^{-1})^{+} (t-t^{-1})^{-} \bar{Q} \non\\
& \qquad + a (p_{q}^{++} + p_{q}^{--}) Q + b (p_{q}^{+} + p_{q}^{-}) 
Q\,,
\end{align}
which is a polynomial in $t$ and $t^{-1}$ iff $a=b=0$. That is, as in 
the rational case, $P$ is 
polynomial iff there is no singular root and no zero root, in which case $\{ u_{1}\,, 
\ldots \,, u_{M}\}$ is admissible. Moreover, $Q_{0,2}$ is polynomial 
iff $a=b=0$. 

\subsubsection{An example}\label{sec:exampleXXZopen}

Let us finally present an explicit example of 
using the Q-system (\ref{QQopenXXZ})-(\ref{icQQopenXXZ}) to compute 
Bethe roots. The 
functions $Q_{0,0}(t)$ and $Q_{1,0}(t)$ are now given by
\begin{verbatim}
Q[0, 0, t_] := (t - t^(-1))^(2L);  
Q[1, 0, t_] := Sum[c[k](t^(2k) + t^(-2k)), {k, 0, M-1}] + t^(2M) + t^(-2M);
\end{verbatim}
and the functions $Q_{1,n}(t)$ and $Q_{0,n}(t)$ are given by
\begin{verbatim}
Q[1, n_, t_] := (Q[1, n-1, t Exp[eta/2]] - 
  Q[1, n-1, t Exp[-(eta/2)]])/(t^2 - t^(-2));
Q[0, n_, t_] := (Q[1, n, t Exp[eta/2]] Q[0, n-1, t Exp[-(eta/2)]] - 
  Q[1, n, t Exp[-(eta/2)]] Q[0, n-1, t Exp[eta/2]])/((t^2 - t^(-2)) Q[1, n-1, t]);
\end{verbatim}
Moreover, we now define $y_{n}(t)$ by
\begin{verbatim}
y[n_, t_] := PolynomialRemainder[Numerator[t^(2L-1) Together[Q[0, n, t]]], 
   Denominator[t^(2L-1) Together[Q[0, n, t]]], t];
\end{verbatim}
We then solve for the coefficients $c_{k}$ and the Bethe roots exactly as in 
Sec. \ref{sec:exampleXXZclosed}..

As before, let us consider the case $N\equiv L=6$ and $M=2$, with
$\eta=\log(2)$.  The above code generates 9 solutions,
as expected (\ref{expectedXXXopen}).  For example, for the first solution, the
corresponding Bethe roots are given by $0.0967267 i\,, 0.385801 i$.

\section{Conclusions}\label{sec:conclusion}

Our main results are $Q$-systems for the closed XXZ
(\ref{eq:QQXXZ})-(\ref{icQQXXZ}), open XXX
(\ref{QQopenXXX})-(\ref{icQQopenXXX}) and open quantum-group-invariant
XXZ (\ref{QQopenXXZ})-(\ref{icQQopenXXZ}) quantum spin chains.
Polynomial solutions of these $Q$-systems can be found efficiently,
which in turn lead directly to the admissible solutions of the
corresponding BA equations.

Numerous applications of these results are possible. In conjunction 
with techniques from algebraic geometry, these $Q$-systems allow the 
exact computation of partition functions for trigonometric vertex 
models and for vertex models with boundaries \cite{Bajnok:2019}.

We restricted here to open spin chains with $SU(2)$ or $U_{q}(SU(2))$ 
symmetry. It would be both interesting and useful to formulate $Q$-systems
for open spin chains with other integrable boundary conditions, as 
well as for integrable models based on $R$-matrices for higher-rank 
algebras and/or higher-dimensional representations.

\section*{Acknowledgments}
ZB and RN are grateful for the warm hospitality extended to them at the 
University of Miami and the Wigner Research Center, respectively. ZB 
was supported in part by NKFIH grant K116505.
The work of EG and JLJ was supported by the European Research Council under the Advanced Grant NuQFT.
RN was supported in part by a Cooper fellowship.

\appendix

%
%
%

\section{Exceptional solutions for open spin chains}\label{sec:exceptional}

\subsection{XXX}\label{subsec:exceptionalXXX}

\subsubsection{$u_{1}=0$}\label{subsec:uzero}

Let us consider a solution $u_{1}\,, \ldots \,, u_{M}$ of the BA 
equations (\ref{BAEopenXXX2}) with one zero Bethe root, say
$u_{1}=0$ (and $u_{2}\,, \ldots \,,
u_{M}$ pairwise distinct and not equal to 0). Since $F_{1}$ in
(\ref{FjXXX}) is not well-defined, we set $u_{1}=\epsilon$ and consider the limit 
$\lim_{\epsilon \rightarrow 0} F_{1}$. It is straightforward to see 
that this limit exists 
and is nonzero. Hence, the corresponding Bethe state in the off-shell 
equation (\ref{offshellXXX}) is {\em not} an 
eigenstate of the transfer matrix, see also \cite{Fendley:1994cz}.

\subsubsection{$u_{1}=\frac{i}{2}$}\label{subsec:usingular}

The equations (\ref{BAEopenXXX1}) evidently have solutions $u_{1}\,,
\ldots \,, u_{M}$ with one 
``singular'' root, say $u_{1}=\frac{i}{2}$ (and $u_{2}\,, \ldots \,,
u_{M}$ pairwise distinct and not equal to $\pm\frac{i}{2}$). However,
the BA equations (\ref{BAEopenXXX2}) do not have such
solutions (recall that the latter equations are not equivalent to
(\ref{BAEopenXXX1}) for this case). Hence, it is not surprising that 
the corresponding Bethe state is not an eigenstate of the transfer 
matrix. Indeed, let us define a renormalized B-operator
\be
\tilde{\B}(u) = \frac{1}{u^{-}} \B(u) \,,
\ee
such that $\lim_{\epsilon \rightarrow 0} \tilde{\B}(\frac{i}{2} + 
\epsilon)$ is finite and non-singular.\footnote{We have checked this 
explicitly for small values of $N$, and we expect that it can
be proved by induction in $N$, similarly to the case of 
the closed chain \cite{Nepomechie:2013mua}.}
Bethe states created with 
this renormalized operator satisfy an off-shell relation similar to 
(\ref{offshellXXX}), except with $F_{j}$ replaced by 
\be
\tilde{F}_{j} = \frac{u^{-}}{u_{j}-\tfrac{i}{2}} F_{j} \,.
\ee
We find that $\lim_{\epsilon \rightarrow 0} \tilde{F}_{1}$
exists and is nonzero for $u_{1} = \pm \tfrac{i}{2} + \epsilon$. 
Hence, the corresponding Bethe state is {\em not} an 
eigenstate of the transfer matrix.

The BA equations (\ref{BAEopenXXX2}) do have solutions with a pair 
of singular Bethe roots, e.g. $u_{1}=\frac{i}{2}$ and
$u_{2}=\pm\frac{i}{2}$, which must be discarded since $|u_{1}|$ and 
$|u_{2}|$ are not distinct.

\subsection{XXZ}\label{subsec:exceptionalXXZ}

For the XXZ case, the coefficients $F_{j}$ of the ``unwanted'' terms are 
given by (\ref{FjXXZ}).
We exclude both $u_{j}=0$ and $u_{j}=\frac{i \pi}{2}$, since 
(similarly to Sec. \ref{subsec:uzero}) the limit
$\lim_{\epsilon \rightarrow 0} F_{1}$ exists and is nonzero
for both $u_{1}=\epsilon$ and $u_{1}=\frac{i \pi}{2} + \epsilon$.
We must also exclude $u_{j} = \pm \frac{\eta}{2}$: similarly to
Sec. \ref{subsec:usingular}, we renormalize the B-operator
\be
\tilde{\B}(u) = \frac{1}{\sinh(2u-\eta)} \B(u) \,,
\ee
so that $\lim_{\epsilon \rightarrow 0} \tilde{\B}(\frac{\eta}{2} + 
\epsilon)$ is finite and non-singular.
The coefficients of the ``unwanted'' terms become
\be
\tilde{F}_{j} = \frac{\sinh(2u-\eta)}{\sinh(2u_{j}-\eta)} F_{j} \,.
\ee 
Then $\lim_{\epsilon \rightarrow 0} \tilde{F}_{1}$ exists and is nonzero
for $u_{1}= \pm \frac{\eta}{2} + \epsilon$.


\begin{thebibliography}{10}

\bibitem{Hagemans:2007}
R.~Hagemans and J.-S. Caux, ``{Deformed strings in the Heisenberg model},''
  {\em J.Phys.A} {\bfseries 40} (2007) 14605,
  \href{http://arxiv.org/abs/0707.2803}{{\ttfamily arXiv:0707.2803
  [cond-mat]}}.

\bibitem{Hao:2013jqa}
W.~Hao, R.~I. Nepomechie, and A.~J. Sommese, ``{Completeness of solutions of
  Bethe's equations},''
  \href{http://dx.doi.org/10.1103/PhysRevE.88.052113}{{\em Phys. Rev.}
  {\bfseries E88} no.~5, (2013) 052113},
\href{http://arxiv.org/abs/1308.4645}{{\ttfamily arXiv:1308.4645 [math-ph]}}.

\bibitem{Marboe:2016yyn}
C.~Marboe and D.~Volin, ``{Fast analytic solver of rational Bethe equations},''
  \href{http://dx.doi.org/10.1088/1751-8121/aa6b88}{{\em J. Phys.} {\bfseries
  A50} no.~20, (2017) 204002},
\href{http://arxiv.org/abs/1608.06504}{{\ttfamily arXiv:1608.06504 [math-ph]}}.

\bibitem{Jiang:2017phk}
Y.~Jiang and Y.~Zhang, ``{Algebraic geometry and Bethe ansatz. Part I. The
  quotient ring for BAE},''
  \href{http://dx.doi.org/10.1007/JHEP03(2018)087}{{\em JHEP} {\bfseries 03}
  (2018) 087},
\href{http://arxiv.org/abs/1710.04693}{{\ttfamily arXiv:1710.04693 [hep-th]}}.

\bibitem{Jacobsen:2018pjt}
J.~L. Jacobsen, Y.~Jiang, and Y.~Zhang, ``{Torus partition function of the
  six-vertex model from algebraic geometry},''
\href{http://arxiv.org/abs/1812.00447}{{\ttfamily arXiv:1812.00447 [hep-th]}}.

\bibitem{Bajnok:2019}
Z.~Bajnok, J.~L. Jacobsen, Y.~Jiang, R.~I. Nepomechie, and Y.~Zhang, ``{in
  progress},''.

\bibitem{Granet:2019}
E.~Granet and J.~L. Jacobsen, ``{On zero-remainder conditions in the Bethe
  ansatz}''.

\bibitem{Pasquier:1989kd}
V.~Pasquier and H.~Saleur, ``{Common Structures Between Finite Systems and
  Conformal Field Theories Through Quantum Groups},''
\href{http://dx.doi.org/10.1016/0550-3213(90)90122-T}{{\em Nucl. Phys.}
  {\bfseries B330} (1990) 523--556}.

\bibitem{Faddeev:1996iy}
L.~D. Faddeev, ``{How algebraic Bethe ansatz works for integrable models},'' in
  {\em Sym\'etries Quantiques (Les Houches Summer School Proceedings vol 64)},
  A.~Connes, K.~Gawedzki, and J.~Zinn-Justin, eds., pp.~149--219.
\newblock North Holland, 1998.
\newblock
\href{http://arxiv.org/abs/hep-th/9605187}{{\ttfamily arXiv:hep-th/9605187
  [hep-th]}}.
\newblock

\bibitem{Izergin:1982hy}
A.~Izergin and V.~Korepin, ``{Pauli principle for one-dimensional bosons and
  the algebraic Bethe ansatz},''
\href{http://dx.doi.org/10.1007/BF00400323}{{\em Lett.Math.Phys.} {\bfseries 6}
  (1982) 283--288}.

\bibitem{Avdeev:1985cx}
L.~Avdeev and A.~Vladimirov, ``{Exceptional solutions of the Bethe ansatz
  equations},''
\href{http://dx.doi.org/10.1007/BF01037864}{{\em Theor. Math. Phys.} {\bfseries
  69} (1987) 1071}.

\bibitem{Nepomechie:2013mua}
R.~I. Nepomechie and C.~Wang, ``{Algebraic Bethe ansatz for singular
  solutions},'' \href{http://dx.doi.org/10.1088/1751-8113/46/32/325002}{{\em J.
  Phys.} {\bfseries A46} (2013) 325002},
\href{http://arxiv.org/abs/1304.7978}{{\ttfamily arXiv:1304.7978 [hep-th]}}.

\bibitem{Pronko:1998xa}
G.~P. Pronko and {\relax Yu}.~G. Stroganov, ``{Bethe equations 'on the wrong
  side of equator'},''
  \href{http://dx.doi.org/10.1088/0305-4470/32/12/007}{{\em J. Phys.}
  {\bfseries A32} (1999) 2333--2340},
\href{http://arxiv.org/abs/hep-th/9808153}{{\ttfamily arXiv:hep-th/9808153
  [hep-th]}}.

\bibitem{Mukhin:2009}
E.~Mukhin, V.~Tarasov, and A.~Varchenko, ``{Bethe algebra of homogeneous XXX
  Heisenberg model has simple spectrum},'' {\em Commun. Math. Phys.} {\bfseries
  288} (2009) 1--42, \href{http://arxiv.org/abs/0706.0688}{{\ttfamily
  arXiv:0706.0688 [math]}}.

\bibitem{Tarasov:2018}
V.~Tarasov, ``{Completeness of the Bethe ansatz for the periodic isotropic
  Heisenberg model},'' \href{http://dx.doi.org/10.1142/S0129055X18400184}{{\em
  Rev. Math. Phys.} {\bfseries 30} (2018) 1840018}.

\bibitem{Nepomechie:2014hma}
R.~I. Nepomechie and C.~Wang, ``{Twisting singular solutions of Bethe's
  equations},'' \href{http://dx.doi.org/10.1088/1751-8113/47/50/505004}{{\em J.
  Phys.} {\bfseries A47} no.~50, (2014) 505004},
\href{http://arxiv.org/abs/1409.7382}{{\ttfamily arXiv:1409.7382 [math-ph]}}.

\bibitem{Thomae:1869}
J.~Thomae, ``{Beitr{\"a}ge zur Theorie der durch die Heinesche Reihe: . . .
  darstellbaren Functionen},'' {\em J. f{\"u}r die reine und angewandte
  Mathematik} {\bfseries 70} (1869) 258--281.

\bibitem{Jackson:1905}
F.~H. Jackson, ``{The basic gamma-function and the elliptic functions},'' {\em
  Proc. Roy. Soc. London} {\bfseries 76} (1905) 127--144.

\bibitem{Baxter:1972wg}
R.~J. Baxter, ``{Eight vertex model in lattice statistics and one-dimensional
  anisotropic Heisenberg chain. I. Some fundamental eigenvectors},''
\href{http://dx.doi.org/10.1016/0003-4916(73)90439-9}{{\em Annals Phys.}
  {\bfseries 76} (1973) 1--24}.

\bibitem{Fabricius:2000yx}
K.~Fabricius and B.~M. McCoy, ``{Bethe's equation is incomplete for the XXZ
  model at roots of unity},''
  \href{http://dx.doi.org/10.1023/A:1010380116927}{{\em J.Statist.Phys.}
  {\bfseries 103} (2001) 647--678},
\href{http://arxiv.org/abs/cond-mat/0009279}{{\ttfamily arXiv:cond-mat/0009279
  [cond-mat.stat-mech]}}.

\bibitem{Baxter:2001sx}
R.~J. Baxter, ``{Completeness of the Bethe ansatz for the six and eight vertex
  models},'' \href{http://dx.doi.org/10.1023/A:1015437118218}{{\em
  J.Statist.Phys.} {\bfseries 108} (2002) 1--48},
\href{http://arxiv.org/abs/cond-mat/0111188}{{\ttfamily arXiv:cond-mat/0111188
  [cond-mat]}}.

\bibitem{Tarasov:2003xz}
V.~O. Tarasov, ``{On Bethe vectors for the XXZ model at roots of unity},''
  \href{http://dx.doi.org/10.1023/B:JOTH.0000049576.42200.77}{{\em J. Math.
  Sciences} {\bfseries 125} (2005) 242--248},
  \href{http://arxiv.org/abs/math/0306032}{{\ttfamily arXiv:math/0306032
  [math.QA]}}.

\bibitem{Gainutdinov:2015vba}
A.~M. Gainutdinov, W.~Hao, R.~I. Nepomechie, and A.~J. Sommese, ``{Counting
  solutions of the Bethe equations of the quantum group invariant open XXZ
  chain at roots of unity},''
  \href{http://dx.doi.org/10.1088/1751-8113/48/49/494003}{{\em J. Phys.}
  {\bfseries A48} no.~49, (2015) 494003},
\href{http://arxiv.org/abs/1505.02104}{{\ttfamily arXiv:1505.02104 [math-ph]}}.

\bibitem{Sklyanin:1988yz}
E.~K. Sklyanin, ``{Boundary Conditions for Integrable Quantum Systems},''
\href{http://dx.doi.org/10.1088/0305-4470/21/10/015}{{\em J. Phys.} {\bfseries
  A21} (1988) 2375}.

\bibitem{Fendley:1994cz}
P.~Fendley and H.~Saleur, ``{Deriving boundary S matrices},''
  \href{http://dx.doi.org/10.1016/0550-3213(94)90369-7}{{\em Nucl.Phys.}
  {\bfseries B428} (1994) 681--693},
\href{http://arxiv.org/abs/hep-th/9402045}{{\ttfamily arXiv:hep-th/9402045
  [hep-th]}}.

\end{thebibliography}

\providecommand{\href}[2]{#2}\begingroup\raggedright\endgroup

\end{document}